\documentclass[%
 reprint,
 amsmath,amssymb,
 aps,nofootinbib
]{revtex4-1}
\usepackage{bigints}
\usepackage{subfig}
\usepackage{bm}
\usepackage{placeins}
\usepackage[bottom]{footmisc}

\usepackage{graphicx}
\graphicspath{ {./figures/} }
\usepackage{dcolumn}
\usepackage{bm}
\usepackage[font={small,it}]{caption}
\usepackage[colorlinks,allcolors=blue]{hyperref}
\usepackage{lipsum}

\newcommand*{\rom}[1]{\expandafter\@slowromancap\romannumeral #1@}

\begin{document}


\title{Stable shapes of three-dimensional vesicles in unconfined and confined Poiseuille flow }

\author{Dhwanit Agarwal and George Biros}
\email{(dhwanit, biros)@oden.utexas.edu}
 \affiliation{Oden Institute of Computational Engineering and Sciences, University of Texas at Austin, TX 78712, USA}

\date{\today}

\begin{abstract}
{We use numerical simulations to  study the dynamics of three dimensional vesicles in unconfined and confined Poiseuille flow. Previous numerical studies have shown that when the fluid viscosity inside and outside the vesicle is same (no viscosity contrast), a transition from asymmetric slippers to symmetric parachutes takes place as viscous forcing or capillary number is increased. At higher viscosity contrast, an outward migration tendency has also been observed in unconfined flow simulations. In this paper, we study how the presence of viscosity contrast and confining walls affect the dynamics of vesicles and present phase diagrams for confined Poiseuille flow with and without viscosity contrast. To our knowledge, this is the first study that provides a phase diagram for 3D vesicles with viscosity contrast in confined Poiseuille flow. The confining walls push the vesicle towards the center while the viscosity contrast has the opposite effect. This interplay leads to important differences in the dynamics like bistability at high capillary numbers.}

\end{abstract}

\pacs{Valid PACS appear here}
\maketitle


\section{\label{sec:level1}Introduction}

Vesicles are closed membranes made of a phospholipid bilayer and serve as a model of nucleus free cells like red blood cells (RBCs). They are filled with fluid and surrounded by fluid. Their high deformability leads to rich shape dynamics in the presence of viscous forcing. Accurate prediction of these shape dynamics when the viscous forcing is generated by a Poiseuille bulk flow, is a fundamental problem since this type of background velocity is predominant in biological flows and microfluidics. For example, Poiseuille flow is used for measuring geometric properties of cells \cite{1}, for understanding the properties of cell suspensions \cite{2}, or for helping in the design of microfluidic devices for sorting cells based on mechanical properties like in lateral displacement devices \cite{3}.

The key parameters that control the shape dynamics are the elastic properties of the membrane, the viscosity contrast (denoted by $\lambda$) between the fluid inside and outside the vesicle (both fluids are typically considered to be Newtonian), the confinement (free vs confined flow, and the confinement ratio defined as the ratio of vesicle diameter to the width of confining channel), and the imposed flow parameters, for example, the velocity magnitude. Regarding the elastic properties, vesicles resist bending but have no resistance in shear or shear rate. A dimensionless parameter called capillary number (denoted by $C_{a}$) measures the ratio of imposed flow strength over the membrane bending energy and is crucial in the study of the shape dynamics. The vesicle membrane is modeled as locally inextensible so there is a surface tension field that enforces this surface inextensibility.  Finally, a key parameter is the reduced volume, $\nu$, of a vesicle, which is the volume of the vesicle over the volume of an equal-area spherical vesicle.  If $\nu=1$ (its maximum value), the vesicle is a sphere that cannot deform and behaves as a rigid particle. For $\nu<1$, the vesicle becomes deformable.  Red blood cells (which are not vesicles because they resist shear) in microcirculation have reduced volume of about 0.7 and viscosity contrast of about five. 

\begin{figure}
    \centering   
    \subfloat[Slipper \label{fig:slipper}]{{\includegraphics[width=2.5cm, height=4.2cm, keepaspectratio]{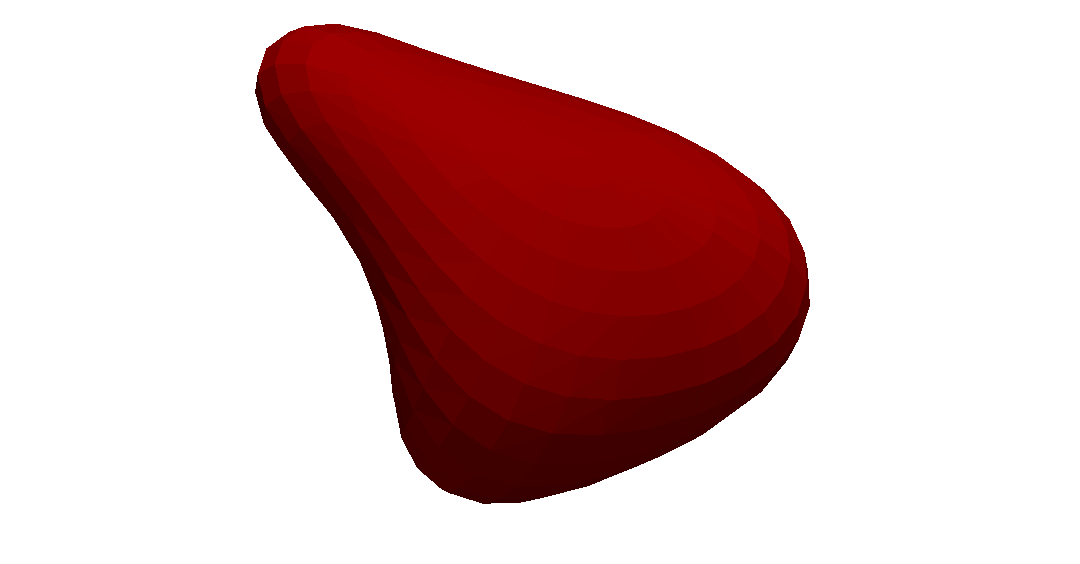} } }
    \subfloat[Croissant \label{fig:croissant}]{{\includegraphics[width=2.5cm, height=1.45cm, keepaspectratio]{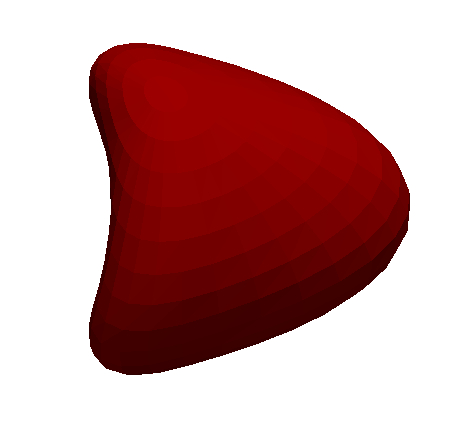} }}
    \subfloat[Parachute\label{fig:parachute}]{{\includegraphics[width=2.5cm, height=4.5cm, keepaspectratio]{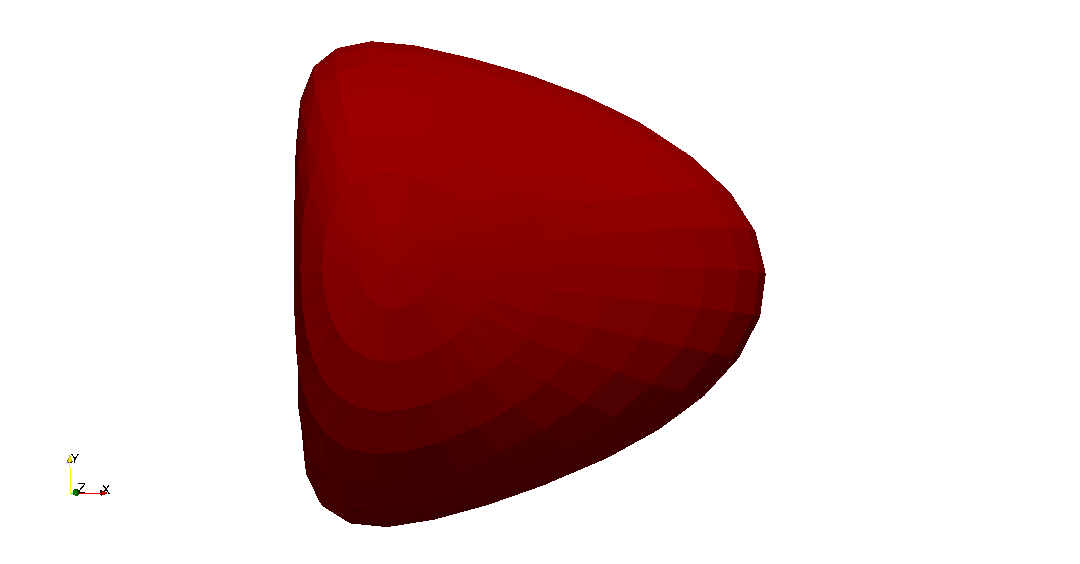} }}
    \caption{Different shapes of vesicles.}
    \label{fig:shapes}
\end{figure}

In Poiseuille flow,  a vesicle evolves into some final shape that typically depends on the elastic parameters of the vesicle and the flow parameters. Two well-known shapes are the \emph{``parachute"} and \emph{``slipper"} (see Fig. \ref{fig:shapes}). Notice that, depending  on the flow conditions, the final state doesn't need to be stationary. It can be oscillatory, for example  \emph{``snaking"} vesicles observed in confined flows. The Poiseuille flow is symmetric; the parachute shape is also symmetric as the center of mass of the vesicle is at the center line, but, surprisingly, the slipper shape is not symmetric and the center of mass is displaced  from the center of the channel. Both shapes were found in one of the earliest experimental studies by Gaehtgens \textit{et al.} \cite{5}.  The existence of slipper shape is mainly attributed to the breakdown of symmetry due to shear gradient in Poiseuille flow.  Kaoui \textit{et al.} \cite{6} confirmed the existence of these stable asymmetric slippers using 2D numerical simulations of a vesicle  in unconfined axisymmetric Poiseuille flow and also presented a phase diagram of parachutes and slippers as a function of velocity and reduced volume. The results established that the parachute occurs at higher velocities while slippers dominate at lower velocities. 2D numerical simulations for vesicles of reduced volume 0.6  in confined Poiseuille flow are presented in  \cite{7} for viscosity contrast one. The simulations revealed a much complex picture as different regimes of confined slipper, unconfined slipper, centered snaking, off centered snaking and parachute appeared depending on the confinement and capillary number. The authors in \cite {10} modeled 2D vesicles in confined Poiseuille flow simulations for viscosity contrast five. The study reveals some important differences in the dynamics when the viscosity contrast is changed from one to five. For example, the authors observed both slipper and parachute shapes at high capillary numbers ($\geq120$) depending on initial position of vesicle.

In 3D, a semi-axisymmetric \emph{croissant} shape (as opposed to fully axisymmetric parachute shape; see Fig. \ref{fig:shapes}) has also been observed in recent simulations by Farutin \textit{et al.} \cite{8} in unconfined Poiseuille flow. That study reported a phase diagram similar to \cite{6} for near spherical vesicles ($\nu \geq 0.9$) with viscosity contrast $\lambda=1$. The authors  observed significant changes in the dynamics at higher viscosity contrast, for example, metastability depending on initial vesicle position and an outward migration tendency.


Coupier \textit{et al.} \cite{28} considered the shapes of 3D vesicles ($\nu \geq 0.91$) in confined Poiseuille flow through experiments for viscosity contrast $\lambda = 1$ and high capillary numbers ($\gtrapprox 15$). Their study revealed a crossover from a parachute to a bullet shape with increasing reduced volume. This point of crossover (the reduced volume above which bullet shape is observed) was observed to be dependent on capillary number for low confinements (confinement ratio $ < 0.5$) while it only depended on confinement ratio at high confinements (confinement ratio $\geq 0.5$). This indicated that the confinement effects dominate the flow strength above 0.5 confinement ratio. Croissant shapes were also observed for rectangular (as opposed to square) channels. 

\emph{Contributions:} In this paper, we build upon the unconfined flow results in \cite{8} and study, through numerical simulations, the dynamics of 3D vesicle of reduced volume $\nu = 0.90$ with and without viscosity contrast. We reproduce (for validation purposes) the phase diagram for unconfined flow for $\lambda =1$, present additional slipper shapes for reduced volume $\nu=0.85$ and study the effects of the presence of viscosity contrast on the dynamics. In the unconfined case, we provide a phase diagram in the parameter space of vesicle initial position and capillary number when $\lambda=5$. For the confined case, we will present phase diagrams in the parameter space of confinement ratio and capillary number for $\lambda \in \{1, 5\}$  and a study of how the confining walls play a pivotal role in determining the dynamics when viscosity contrast is present. We will see how the wall effects and the outward migration due to viscosity contrast lead to coexistence of both slipper and parachute shapes in the confined case. Our results could explain the experiments with RBCs in which slipper shapes are observed at high capillary numbers \cite{29, 30}. We will also see that at high confinement ratio ($\geq 0.5$), the wall effects dominate causing the vesicle to remain centered with mostly axisymmetric shapes. We also observe two new equilibrium shapes, \emph{``bean"} and \emph{``bell"}, in the confined case.  

\emph{Limitations:} It is important to stress here that the vesicles don't have any shear resistance. This makes them different from RBCs which resist shear. Thus, vesicles serve only as a simplistic model of these cells and can have different dynamics compared to the RBCs. We would also like to specify that for smaller reduced volumes ($\nu\leq 0.85$), our numerical scheme is currently unable to resolve the vesicle shapes at high $C_{a}$ ($>10$) due to large deformations. 

This paper is organized as follows. In Sec. \ref{sec:method}, we present
the problem statement and methodology for both unconfined and confined flows. In Sec. \ref{sec:parameters}, we formulate the relevant parameters in both confined and unconfined cases, tabulate the self convergence results and verify the correctness of our code by comparing our results and shapes with previous literature. In Sec. \ref{sec:phaseunconfined}, we present
 the results regarding different steady state shapes and behavior of vesicles including the phase diagram for unconfined flow. In Section \ref{sec:phaseconfined}, we discuss the results for a vesicle in a confined Poiseuille flow in detail. 
 In Sec. \ref{sec:conc}, we present the conclusion and further ideas to be explored.

\section{\label{sec:method}Problem Formulation and Methodology}

In this section, we state the flow problem and give its boundary integral formulation.
The detailed derivation of this formulation is given in \cite{21}.  Table \ref{symbols} summarizes the notation used in the paper.

\begin{table}[h!]
 \begin{tabular}{|c c|}
 \hline
 Symbol & Definition\\ [0.5ex] 
 \hline\hline
 $\mathbb{S}^{2}$ & Unit Sphere \\
 \hline
  $p$ & Degree of spherical harmonic expansion  \\ 
 \hline
 $\gamma$ & Boundary of vesicle \\
 \hline
 $\Gamma$ & Fixed rigid boundary  \\
 \hline
 $\mathbf{S}_{\gamma}$ &  The single-layer Stokes operator 
over  $\gamma$ \\
 \hline
 $\mathbf{D}_{\gamma}$ &  The double-layer Stokes operator
over  $\gamma$ \\
 \hline
 $\boldsymbol{u}$ & Velocity \\
 \hline
 $\boldsymbol{u}_{\infty}$ & Background velocity \\
 \hline
  $\mathbf{f}$ & Interfacial force\\
 \hline
  $P$ & Pressure \\
 \hline
  $t$ & Time \\
 \hline
 $\mathbf{n}$ & Outward unit normal\\
 \hline
 $\mu_\mathrm{i} $ & Viscosity of fluid in vesicle\\
 \hline
 $\mu_\mathrm{e}$ & Viscosity of ambient fluid\\
 \hline
 $\lambda$ & Viscosity contrast = $\mu_\mathrm{i} / \mu_\mathrm{e}$\\
 \hline
 $\sigma$ & Tension\\
 \hline
 $H$ & Mean curvature of vesicle \\
 \hline
 $K$ & Gaussian curvature of vesicle \\
 \hline
 $\kappa_{b}$ & Bending modulus of vesicle membrane \\
 \hline
 $\bm{\eta}$ & Double layer density on $\Gamma$\\
 \hline
 $\omega$ & Volume enclosed by $\gamma$\\
 \hline
 $\Omega$ & Volume of interest\\
 \hline
 $R_{0}$ & Radius of vesicle\\
 \hline
 $\nu$ & Reduced volume of vesicle\\
 \hline
  $C_{a}$ & Capillary number\\
 \hline
 $C_{n}$ & Confinement ratio\\
 \hline
 \end{tabular}
 \captionof{table}{Index of different symbols}\label{symbols}
\end{table}

\begin{figure}%
    \centering
    \subfloat[Unconfined flow \label{fig:schematic_uncon}]{{\includegraphics[width=3.1cm]{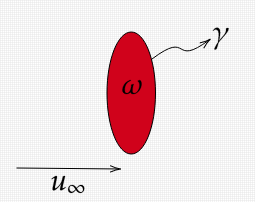} } }%
    \qquad
    \subfloat[Confined flow \label{fig:schematic_con}]{{\includegraphics[width=4.5cm]{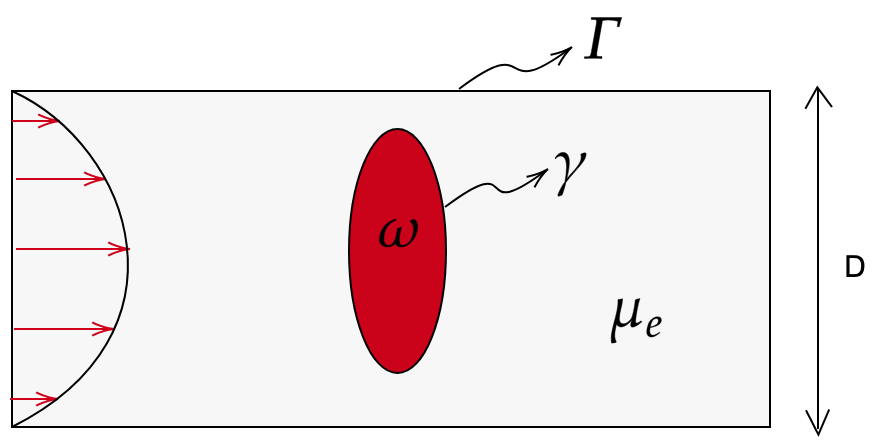} }}%
    \caption{(a) Schematic of the domain $\Omega$ for unconfined flow. $\omega$ denotes the red region enclosed by vesicle membrane $\gamma$ filled with fluid of viscosity $\mu_{i}$.  Grey region is filled with fluid of viscosity $\mu_{e}$. $u_{\infty}$ is the imposed background velocity. (b) Schematic of the domain $\Omega$ for confined flow (side view).  $\Gamma$ is the fixed rigid enclosing boundary, which models a channel of circular cross-section with axis  parallel to x-axis. $D$ is the diameter of the circular cross section. Parabolic flow imposed is shown by red arrows. Grey region is filled with fluid of viscosity $\mu_{e}$. $\omega$ denotes the red region enclosed by vesicle membrane $\gamma$ filled with fluid of viscosity $\mu_{i}$.}%
    \label{fig:schematic}%
\end{figure}

\subsection{\label{sec:unconfinedmodel} Unconfined flow formulation}

The formulation for unconfined flow of vesicles is the same as in \cite{22}. In Fig. \ref{fig:schematic_uncon}, we show the geometric setup for our simulations. In this setup, the fluid flow is governed by the Stokes equation due to negligible effect of inertial forces. The PDE formulation of the flow is as follows:

\begin{align} 
-\mu(\boldsymbol{x}) \Delta \boldsymbol{u}(\boldsymbol{x}) + \nabla P(\boldsymbol{x}) = 0   \textbf{    } \forall \mathbf{x} \in \mathbb{R}^{3}\backslash \gamma  \label{eq1},\\
div (\boldsymbol{u}(\boldsymbol{x}) )= 0  \textbf{     } \forall \mathbf{x} \in \mathbb{R}^{3}\backslash \gamma  \label{eq2},\\
[[ -P\mathbf{n} + (\nabla \boldsymbol{u} + \nabla \boldsymbol{u}^{T})\mathbf{n}]] = \mathbf{f}  \textit{ on  }  \gamma \label{eqjump},\\
\frac{\partial \mathbf{X}}{\partial t} = \boldsymbol{u}(\mathbf{X})  \textbf{     }  \forall  \mathbf{X} \in \gamma \label{eq3}, \\
 \boldsymbol{u}(\mathbf{x})  \rightarrow \boldsymbol{u}_{\infty} (\mathbf{x}) \textit{ as   }  \mathbf{||x||} \rightarrow \infty \label{eqinf}, 
\end{align}
where $\gamma$ is the vesicle membrane, $\mathbf{u}(\mathbf{x})$ is the velocity of  the fluid and  $P(\mathbf{x})$ is the pressure. The viscosity $\mu$ is given by 
\[ \mu(\boldsymbol{x}) = \begin{cases} 
      \mu_\mathrm{i} & \text{ if } \boldsymbol{x} \in \omega, \\
       \mu_\mathrm{e} & \text{ if } \boldsymbol{x} \in \mathbb{R}^{3} \backslash \omega. \\
     \end{cases}
\]
$[[l]]$ denotes the jump of quantity $l$ across the vesicle membrane and $\mathbf{n}$ is the outward unit normal to the membrane. Equation (\ref{eqjump}) is the balance of momentum on membrane, which requires the surface traction jump to be equal to the total force (denoted by $\mathbf{f}$) exerted by the interface onto fluid. Equations (\ref{eq3}-\ref{eqinf}) enforce no-slip boundary condition on vesicle membrane and set the far field velocity to be the background velocity. We use $\mathbf{X}$ to denote a point on the vesicle membrane $\gamma$ while $\mathbf{x}$ to denote a point in $\mathbb{R}^{3}\backslash \gamma$.
 
 The local inextensibility of vesicle membrane is mathematically equivalent to requiring that the surface divergence of velocity should vanish on vesicle membrane, i.e., 
 \begin{align}
div_{\gamma} (\boldsymbol{u}(\boldsymbol{X}) )= 0  \textbf{     } \forall \mathbf{X} \in  \gamma  \label{eqsurfdiv}.
\end{align}
Now let us discuss in detail the elastic force $\mathbf{f}$ due to the vesicle membrane elasticity. It comprises a bending and a tension component, the latter being a Lagrange multiplier that enforces the local inextensibility. We denote the bending component by $\mathbf{f}_{b}$ and tension component by $\mathbf{f}_{\sigma}$, so we write 
\begin{align}
\mathbf{f} = \mathbf{f}_{b} + \mathbf{f_{\sigma}}.
\end{align}
The expressions for these components are (please refer to \cite{23, 24} for details): 
 \begin{align}
 \mathbf{f_{b}}(\mathbf{X}) = -\kappa _{b} [\Delta_{\gamma}H + 2H(H^{2} - K))] \mathbf{n}, \\
 \mathbf{f_{\sigma}}(\mathbf{X}) = \sigma \Delta_{\gamma}\mathbf{X} + \nabla_{\gamma} \sigma,
 \end{align} 
where $\kappa_{b}$ is membrane's bending modulus, $H$ and $K$  are the mean and Gaussian curvature respectively, $\sigma$ is the tension at the membrane point $\mathbf{X}$. 

Following \cite{21, 25}, we can rewrite these equations in integral form for $\mathbf{X} \in \gamma$ as follows : 
\begin{align}
\alpha \boldsymbol{u}(\mathbf{X}) =  \boldsymbol{u}_{\infty} (\mathbf{X})  + \mathbf{S}_{\gamma}[\mathbf{f_{b}} + \mathbf{f_{\sigma}}] (\mathbf{X}) + \mathbf{D}_{\gamma}[\boldsymbol{u}] (\mathbf{X}) \label{syseq1},\\
div_{\gamma}(\boldsymbol{u} (\mathbf{X})) = 0 \label{syseq2},\\
\frac{\partial \mathbf{X}}{\partial t} = \boldsymbol{u} ({\mathbf{X}}) \label{syseq3},
\end{align}
where $\alpha := (1 + \lambda)/ 2$. The single layer convolution integral is defined as  $\mathbf{S}_{\gamma}[f] (\mathbf{x}) := \bigints_{\gamma} S_{0}(\mathbf{x}, \mathbf{y}) \mathbf{f(\mathbf{y})} d\gamma $, with

\begin{align*}
S_{0}(\mathbf{x}, \mathbf{y} ) = \frac{1}{8 \pi \mu} \frac{1}{||\mathbf{r}||} (I + \frac{\mathbf{r} \otimes \mathbf{r}}{||\mathbf{r}||^{2}}),
\end{align*}
where $\mathbf{r}:= \mathbf{x} - \mathbf{y}$, $I$ is the identity operator, $\otimes$ is tensor product and $||\cdot||$ is the Euclidean norm.  The double layer convolution integral is defined as $\mathbf{D}_{\gamma}[f] (\mathbf{x}) := \bigints_{\gamma} D_{0}(\mathbf{x}, \mathbf{y}) \mathbf{f(\mathbf{y})} d\gamma $, with
  
  \begin{align*}
  D_{0}(\mathbf{x}, \mathbf{y} ) = \frac{-3(1- \lambda)}{4 \pi}((\mathbf{r}. \mathbf{n})\frac{\mathbf{r} \otimes \mathbf{r}}{||\mathbf{r}||^{5}}).
  \end{align*}

\textit{Discretization:} We use spherical harmonics discretization for $\mathbf{X}$ and functions defined on $\gamma$. The singular quadratures described in \cite{22} are used to evaluate the integrals. The system of equations (\ref{syseq1})--(\ref{syseq3}) is then solved using a semi-implicit scheme \cite{22} for the velocity $\mathbf{u}$ and tension $\sigma$. Vesicle position, $\mathbf{X}$, is then updated as $\mathbf{X}_{new} = \mathbf{u}\Delta t + \mathbf{X}_{old}$, where $\Delta t$ is the time step.


\subsection{\label{sec:confinedmodel} Confined flow formulation}
We define vesicle radius, denoted by $R_{0}$, to be the radius of the sphere that has the same volume as the vesicle. We set $R_{0}=2$ in our simulations. To model the flow of a vesicle in confined Poiseuille flow, we create a channel with length much larger than the vesicle radius and a circular cross-section. In our simulations, we set the length of channel to be eight times the vesicle radius $R_{0}$. The axis of the channel is parallel to $x$-axis and vesicle starts slightly displaced in the $y$-direction from the axis of the channel. Refer to Fig. \ref{fig:schematic_con} for a general representation of the setup. The boundary integral formulation that accounts for confinement is the 3D extension of the formulation discussed in \cite{26}. To account for the confinement, we add the vesicle-wall interaction term to the RHS of the equation (\ref{syseq1}) and append one more equation (\ref{consyseq3}) for the calculation of the unknown double layer density $\bm{\eta}$ on the fixed rigid boundary $\Gamma$. The formulation becomes: 
 \begin{align}
\alpha \boldsymbol{u}(\mathbf{X}) &=   \mathbf{S}_{\gamma}[\mathbf{f_{b}} + \mathbf{f_{\sigma}}] (\mathbf{X}) + \mathbf{D}_{\gamma}[\boldsymbol{u}] (\mathbf{X})  +  \mathbf{D}_{\Gamma}[\bm{\eta}](\mathbf{X}) ,  \textit{           }    \label{consyseq1}  \\
 div_{\gamma}(\boldsymbol{u} (\mathbf{X})) &= 0 \label{consyseq2} \textit{      } \forall \mathbf{X} \in  \gamma ,  \\
\mathbf{U}(\mathbf{x}) &= -\frac{1}{2}\bm{\eta}(\mathbf{x}) + \mathbf{S}_{\gamma}[\mathbf{f_{b}} + \mathbf{f_{\sigma}}] (\mathbf{x}) + \mathbf{D}_{\gamma}[\boldsymbol{u}] (\mathbf{x}) \notag \\
&\phantom{{}=1} +  \mathbf{D}_{\Gamma}[\bm{\eta}](\mathbf{x}) + \mathbf{N}_{0}[\bm{\eta}](\mathbf{x})    \textit{            } \forall \mathbf{x} \in \Gamma,  \label{consyseq3}\\
\frac{\partial \mathbf{X}}{\partial t} &= \boldsymbol{u} ({\mathbf{X}})   \textit{            } \forall \mathbf{X} \in \gamma,   \label{consyseq4}
\end{align}  
where $\mathbf{N}_{0}[\mathbf{\bm{\eta}}](\mathbf{x})  = \mathbf{n}(\mathbf{x})\bigints_{\Gamma} \left( \mathbf{n}(\mathbf{y})\cdot \bm{\eta}(\mathbf{y})\right) ds(\mathbf{y})$ and $\mathbf{U}(\mathbf{x})$ is the given velocity of rigid enclosing boundary at $\mathbf{x} \in \Gamma$. We solve the system of equations (\ref{consyseq1}-\ref{consyseq2}) for $\mathbf{u}$ and $\sigma$ as in the unconfined case. We then use the obtained $\mathbf{u}$ and $\sigma$ in (\ref{consyseq3}) to solve for double layer density $\bm{\eta}$ on $\Gamma$. Finally, equation (\ref{consyseq4}) is discretized as  $\mathbf{X}_{new} = \mathbf{u}\Delta t + \mathbf{X}_{old}$ to solve for new vesicle position $\mathbf{X}_{new}$. To avoid the effect of finite length of the channel, after each time step we translate the vesicle so that the $x$-coordinate of the center of vesicle coincides with the $x$-coordinate of the center of the channel. Each simulation typically takes about 10000 time steps and 15 hours of wall clock time.

\section{\label{sec:parameters} Simulation setup and parameters}

In this section, we describe the simulation setup and list the relevant input and output parameters which we monitor to study the dynamics of vesicles in both unconfined and confined case.   

\subsection{\label{sec:unconfinedsim} Unconfined flow parameters}
A vesicle is characterized  by its reduced volume $\nu$, which is defined as the ratio of the volume of vesicle to the sphere with the same area as the vesicle. It is given by 
 \begin{align*}
 \nu:= 6 \pi^{1/2} V A^{-3/2},
 \end{align*}
where $V$ and $A$ are the volume and surface area of the vesicle respectively. The imposed background fluid flow $\boldsymbol{u}_{\infty} = (v_{x}, 0, 0)$  is an axisymmetric Poiseuille profile given by  
\begin{align}
 v_{x} = \alpha (D^{2}/4 - y^{2} - z^{2})
 \end{align}
 in Cartesian coordinates, where $D$ is the diameter of the Poiseuille flow and $\alpha$ is the curvature of the flow. Vesicle starts slightly displaced in the $y$-direction from the centerline of Poiseuille flow. We use the dimensionless capillary number for Poiseuille flow given by 
 \begin{align}
 C_{a} := \frac{\alpha R_{0}^{4} \mu_\mathrm{e}}{\kappa_{b}}, \label{ca}
 \end{align}
 where $\kappa_{b}$ is the bending modulus of the vesicle as specified in the Sec. \ref{sec:unconfinedmodel}. $C_{a}$ measures the flow strength over the bending
energy of the membrane. We take $R_{0}=2$, $D= 20R_{0}$, $\mu_{e}=1$ and $\kappa_{b}=1$ in our simulations\footnote{If the unit of length is micrometers ($\mu m $), unit of mass is microgram ($\mu g$) and the time is in seconds ($s$), then $R_{0} = 2\mu m$, $\kappa_b = 10^{-21} \text{ Joules}$, $\mu_{e} = 10^{-3} \text{ Pascal. second}$. Then if the flow curvature is $\alpha=1$ in our simulation, it corresponds to a Poiseuille flow with a  maximum velocity of $400\mu m /s$.} and vary the flow curvature $\alpha$ to vary the capillary number $C_{a}$. Although the flow is unconfined, we can measure degree of confinement using dimensionless confinement ratio ($C_{n}$), defined as $C_{n} := \frac{2R_{0}}{D}$. Since $D$ and $R_{0}$ are fixed, $C_{n}=0.1$ in our unconfined flow simulations. The viscosity contrast $\lambda:= \mu_\mathrm{i}/ \mu_\mathrm{e}$ is the ratio of the viscosity of the fluid inside the vesicle to viscosity of the outside fluid and is crucial in determining vesicle dynamics. We study the vesicle dynamics for a range of capillary numbers and reduced volumes with $\lambda=1$ and $\lambda=5$.

 \subsection{\label{sec:confinedsim} Confined flow parameters}
In our simulations of confined flow, the diameter of the channel is denoted by $D$. The velocity of the rigid boundary $\Gamma$ is denoted by $\mathbf{U(\mathbf{x})}$ (refer to Fig. \ref{fig:schematic_con}). We impose the Poiseuille velocity profile with diameter of the flow equal to $D$. To do this, we set  the velocity of the rigid boundary 
 \begin{align*}
 \mathbf{U(x)} = (v_{x}, 0, 0), \\
 \text{ where } \mathbf{x}=(x, y, z) \text{ and } v_{x} = \alpha(D^{2}/4 - y^{2} - z^{2}),
 \end{align*}
 in equation (\ref{consyseq3}). We use capillary number $C_{a}$, reduced volume $\nu$ and viscosity contrast $\lambda$ as defined earlier. We also use a dimensionless parameter called confinement ratio ($C_{n}$), given by 
  \begin{align*}
  C_{n} = \frac{2R_{0}}{D},
 \end{align*}
 to characterize the extent of confinement. The higher the $C_{n}$ is, the closer the bounding walls are to the vesicle. We study the vesicle dynamics for a range of capillary numbers and confinement ratios. 
 
\subsection{\label{sec:convergence} Steady state and convergence results}
To determine if a steady state is reached, we monitor the lateral displacement of vesicle center ($Y_{g}$), bending energy of the vesicle ($E_{b} := \bigints_{\gamma} \frac{1}{2} \kappa_{b} H^{2} d \gamma $ ) and its volume moments tensor ($\mathbf{I} := \bigints_{\omega} (||\mathbf{\hat{r}}||^{2} I - \mathbf{\hat{r}} \otimes \mathbf{\hat{r}}) dV$, where $\mathbf{\hat{r}} = \mathbf{r} - \mathbf{r_{0}}$, $\mathbf{r_{0}}$ is the center of the vesicle and $I$ is the identity tensor). We say that the vesicle has reached a steady state when these observables reach a steady state or their dynamics become nearly periodic.  Such oscillatory behavior is actually the typical scenario in our simulations. 

To verify the correctness of our code, we report the self-convergence results in a variety of different settings. First, we consider the unconfined setting for fixed physical parameter values of $\nu=0.90, C_{a} = 5$ and $\lambda =1$. In this scenario, we compare the position of the vesicle center for spherical harmonics discretization of order $p=6,12, 24$ and $48$ after several thousand time steps. We regard the simulation with $p=48$ as the ground truth and compute the relative error in the position of vesicle's center as a function of spherical harmonics order $p$. We repeat this for confined flow with $\nu=0.90, C_{n}=0.5, C_{a} = 5$ and $\lambda=1$. The results are summarized in Table \ref{conv}.

\begin{table}[h!]
 \begin{tabular}{|c | c | c|}
 \hline
 $p$ & Rel. error (unconfined flow)  & Rel. error(confined flow) \\ [0.5ex] 
 \hline \hline
 $6$ & $4.1 \times 10^{-3}$ & $6 \times 10^{-2}$ \\
 
 $12$ & $6 \times 10^{-5}$ & $4 \times 10^{-3} $ \\
 
 $24$ & $1.2 \times 10^{-5}$  & $1.5 \times 10^{-4}$\\
 \hline
 \end{tabular}
  
 \captionof{table}{ Self convergence results for unconfined and confined flow. $p$ is the order of spherical harmonics discretization. We regard $p=48$ as the ground truth and calculate error relative to it.}\label{conv}
\end{table}

\begin{figure}%
    \centering   
    \subfloat[Side view of slipper]{{\includegraphics[width=3.8cm, height=4cm,  keepaspectratio]{ves278_paper1.png} } }
    \subfloat[Rear view of slipper]{{\includegraphics[width=3.8cm, height=4cm, keepaspectratio]{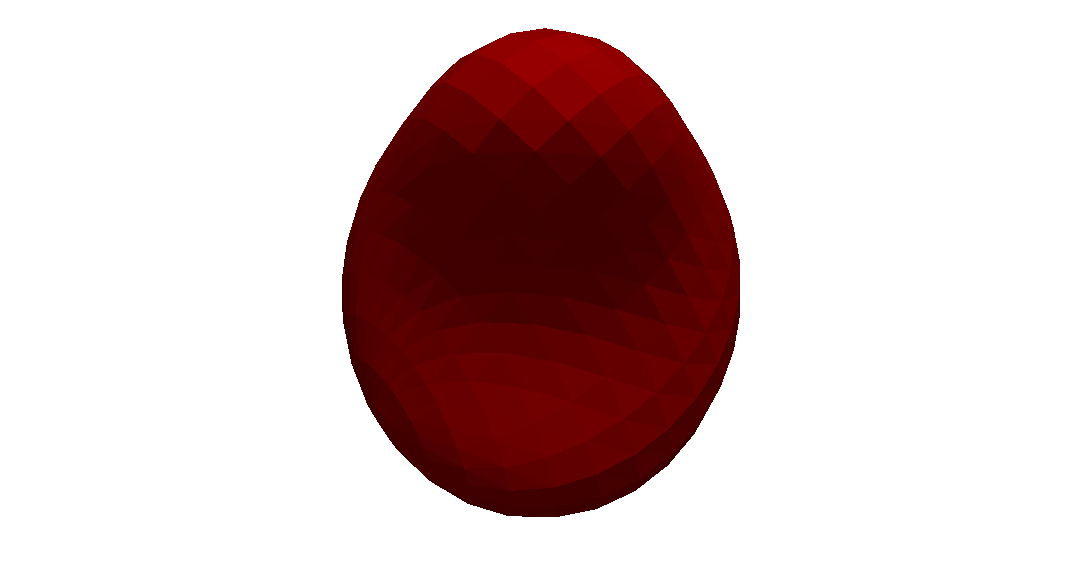} }}
    \caption{Unconfined flow with $\nu =0.90, C_{a}=4, \lambda=1$. Tank treading slipper. \label{fig:slipper_test} }
    \subfloat[Side view of croissant]{{\includegraphics[width=3.8cm]{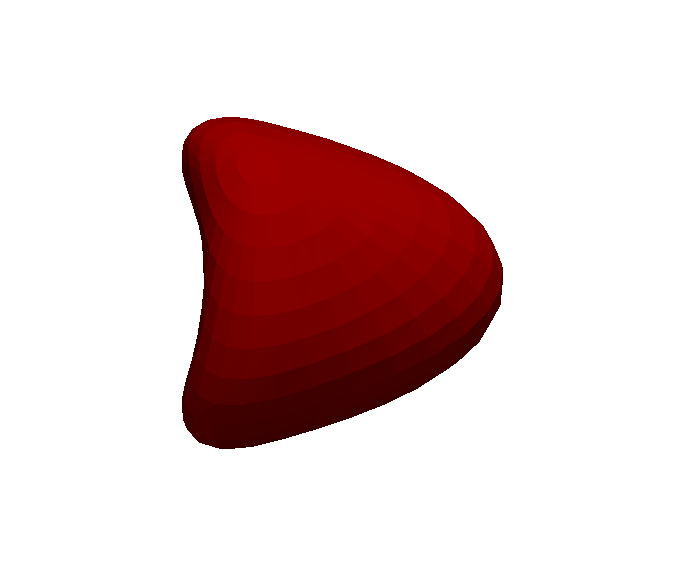} }}
    \subfloat[Rear view of croissant]{{\includegraphics[width=3.8cm]{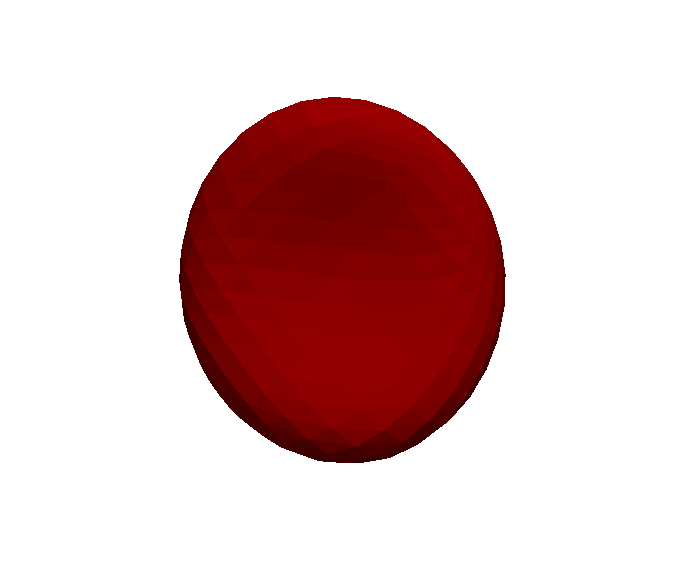} }}
    \caption{Unconfined flow with $\nu =0.90, C_{a}=14, \lambda=1$. Semi-axisymmetric croissant shape. No tank treading. . \label{fig:croissant_test}}
      \subfloat[Side view of parachute]{{\includegraphics[width=4.0cm]{ves249_paper2.png} }}
    \subfloat[Rear view of parachute]{{\includegraphics[width=3.8cm]{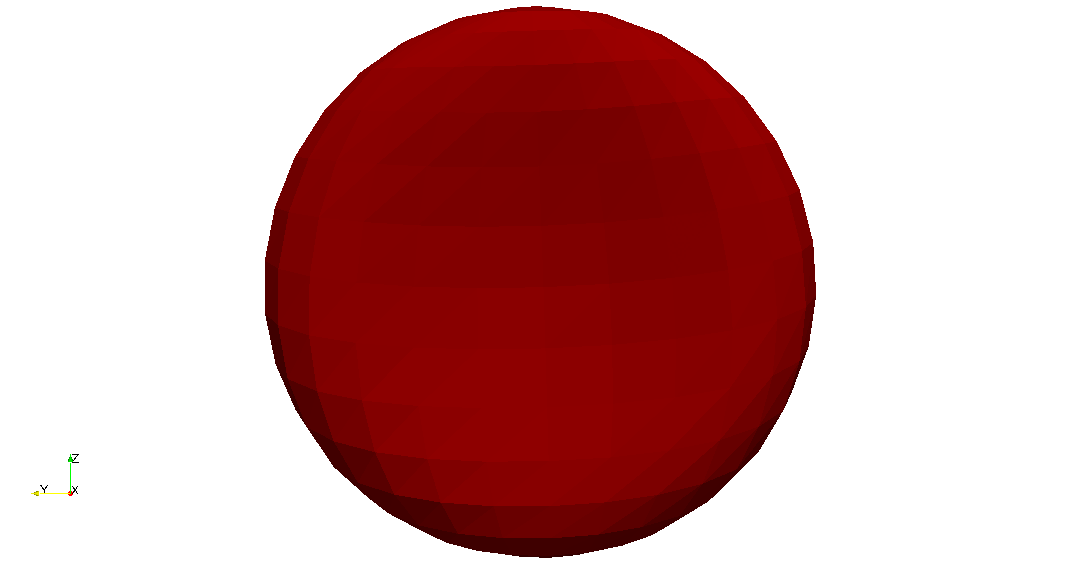} }}
    \caption{Unconfined flow with $\nu =0.90, C_{a}=28, \lambda=1$. Centered axisymmetric parachute shape. No tank treading. \label{fig:parachute_test}}
      \subfloat[Side view of bullet]{{\includegraphics[width=3.8cm]{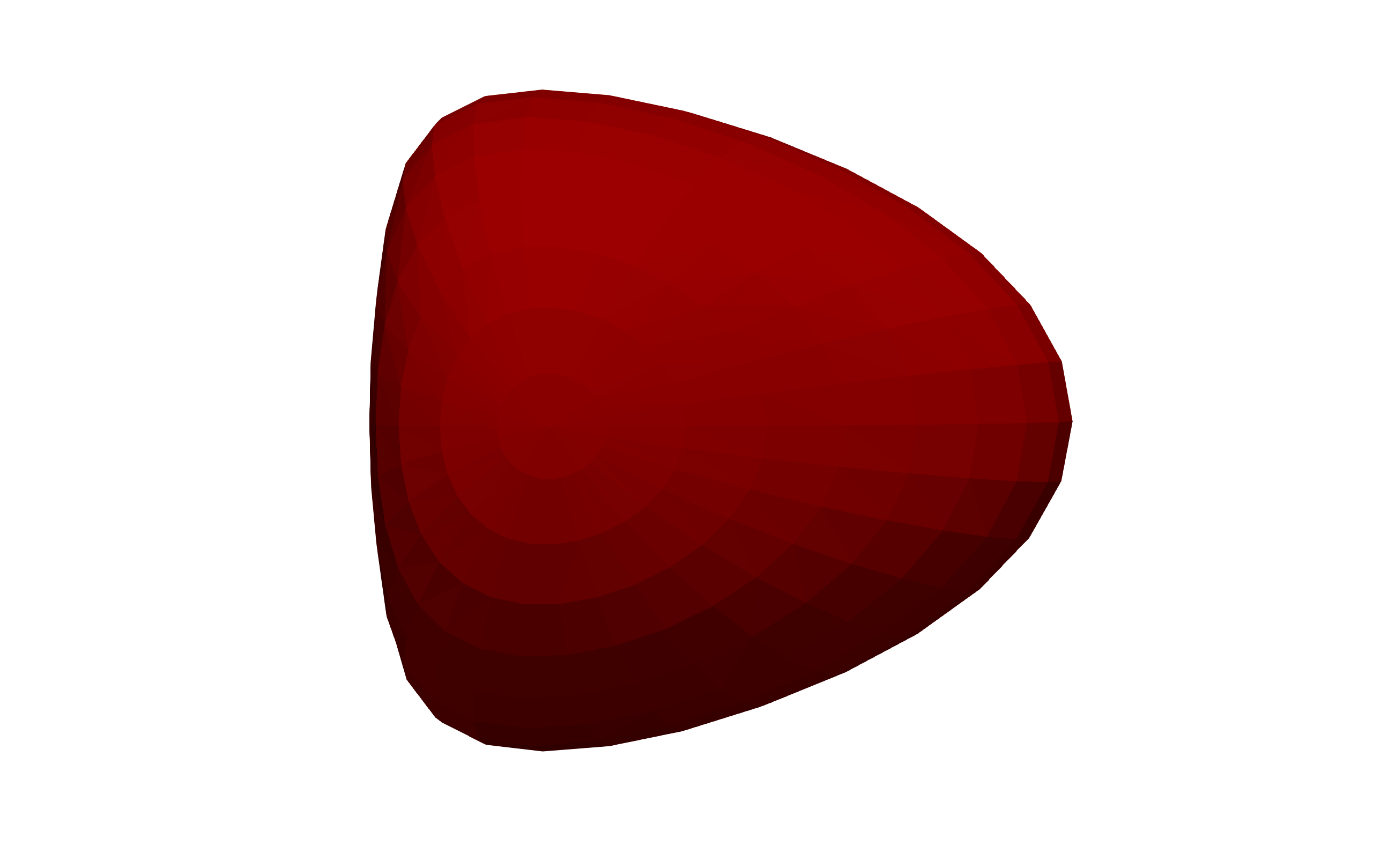} }}
    \subfloat[Rear view of bullet]{{\includegraphics[width=3.8cm]{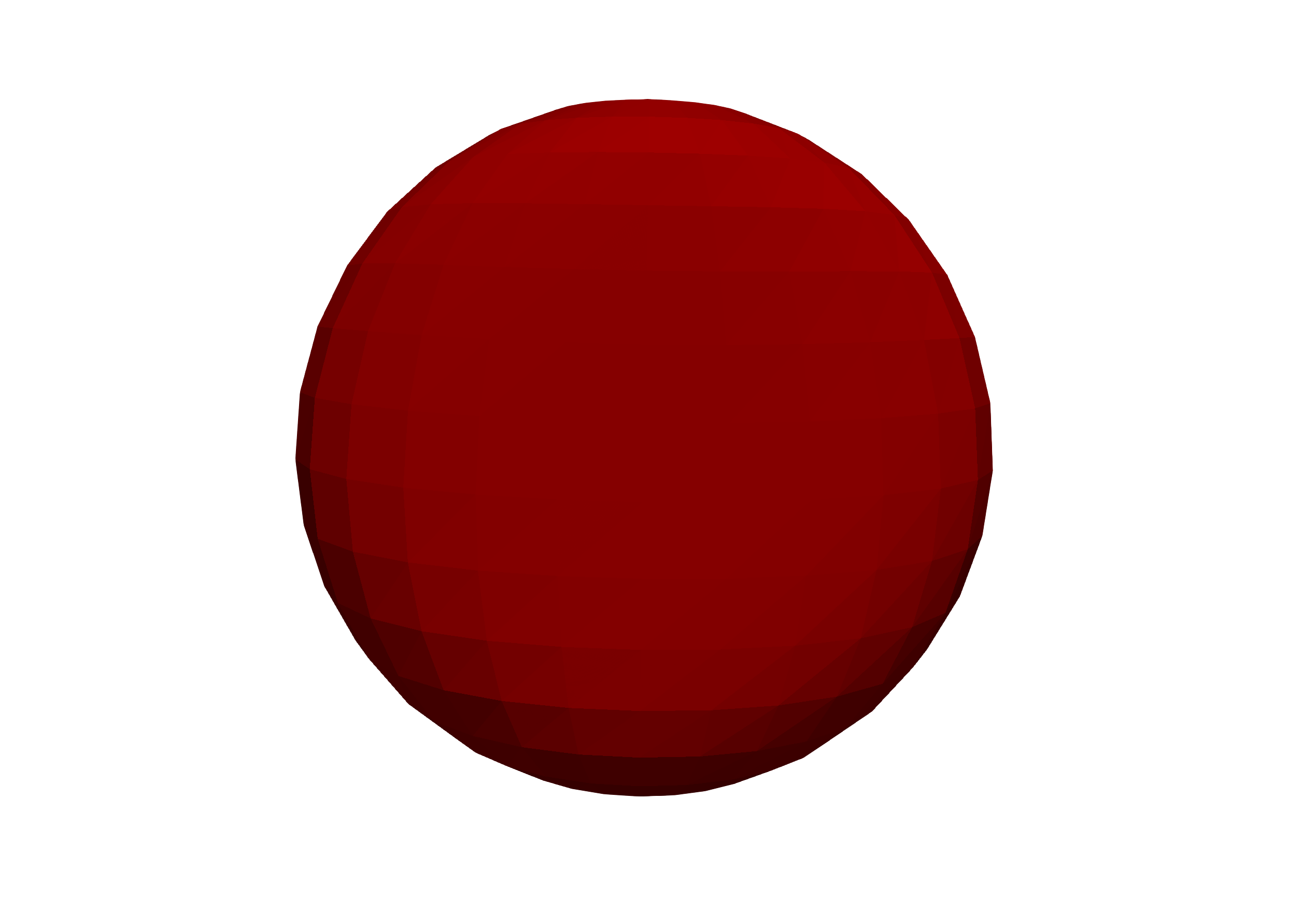} }}
    \caption{Unconfined flow with $\nu =0.96, C_{a}=500, \lambda=1$. Centered axisymmetric bullet shape. No tank treading. \label{fig:bullet_test}}
    \includegraphics[width=4.6cm]{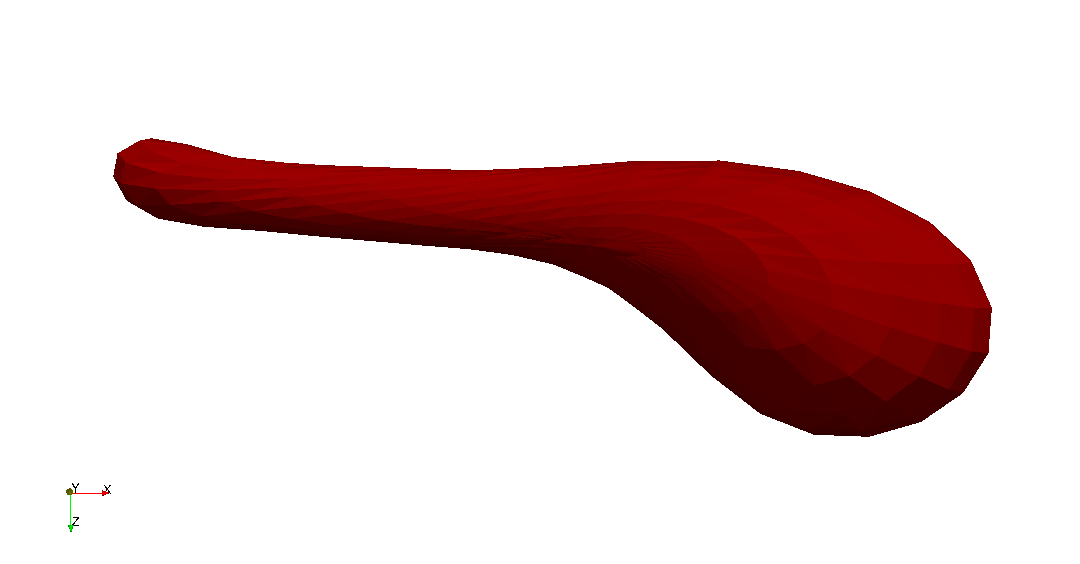} 
    \caption{Unconfined flow simulation $\nu =0.65, C_{a}=1.8, \lambda=1$. Tank treading slipper shape. }
    \label{fig:slipperdeflated}

\end{figure}

\begin{figure*}%
    \centering
    \subfloat[ Vesicle center ($Y_{g}$) scaled with $R_{0}$]{{\includegraphics[width=5.8cm, height=4.1cm]{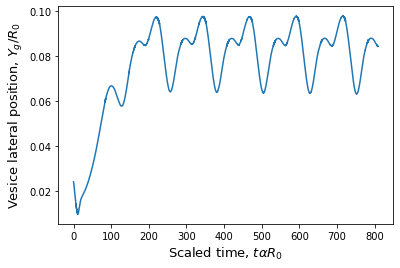} }}
    \subfloat[Bending energy of vesicle with time ]{{\includegraphics[width=5.8cm, height=4.6cm]{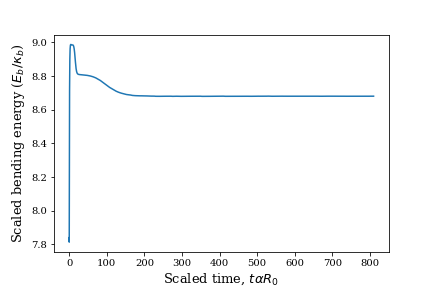} }}
    \subfloat[Vesicle moment about flow axis with time ]{{\includegraphics[width=5.8cm,  height=4.6cm]{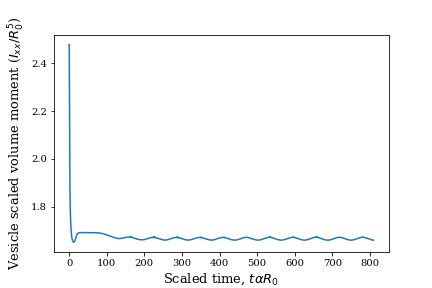} }}
    \caption{Periodic oscillations in vesicle position and moment  in unconfined flow simulation with $\nu =0.90, \lambda=1, C_{a}=4$. Bending energy remains constant after some time. Equilibrium shape is a slipper.}%
    \label{fig:posmombeslipper}%
\end{figure*}

To further verify the correctness, we present the steady state shapes of our simulations of unconfined flow for $\nu = 0.9, \lambda=1$ with $C_{a}=4, 14 $ and $28$ in Fig. \ref{fig:slipper_test}, Fig. \ref{fig:croissant_test} and Fig. \ref{fig:parachute_test} respectively. In these simulations, we obtained three shapes, namely 1) tank treading off centered slipper (called TT slipper), 2) non-tank treading slightly off centered croissant and 3) non-tank treading centered parachute. These shapes and results are consistent with the numerical and experimental results presented in \cite{8, 28} and we view them as an additional validation of our code (see Fig. \ref{fig:phase_uncon}).  In the case when we obtain a slipper shape, i.e., when  $C_{a} = 4$, we obtain periodic oscillations of vesicle position while bending energy remains constant (see Fig. \ref{fig:posmombeslipper}). 
Axisymmetric bullet shape with flat rear (as opposed to concave rear in parachute) is obtained for $\nu=0.96, C_{a}=500, \lambda=1$ shown in Fig. \ref{fig:bullet_test}. We also present the steady state shape obtained for unconfined flow simulation with a vesicle of reduced volume $\nu =0.65$ for $\lambda=1$ and $C_{a} = 1.8$ in Fig. \ref{fig:slipperdeflated}. The shape obtained is similar to the one presented in \cite{20}. We observe periodic oscillations of slipper shape about a mean position in this case as well.  

\section{\label{sec:phaseunconfined}Results for unconfined flow}

\subsection{Viscosity contrast $\lambda=1$}
\begin{figure*}
    \centering
    \subfloat[ A phase diagram for unconfined flow with $\lambda=1$. Shows equilibrium shapes in different regions of vesicle reduced volume $\nu$ and capillary number $C_{a}$. Green denotes slipper, blue denotes croissant, red denotes parachute and magenta denotes bullet. Triangles, squares, diamond and circles denote our simulations. Hexagrams denote experimental low confinement results in \cite{28} and $+$ denote simulation results in \cite{8}. The unresolved regime is the range of parameter values for which our code was unable to resolve the shapes. Dashed black curves are a guide to the eye. \label{fig:phase_uncon}]{{\includegraphics[width=10 cm, height=7cm]{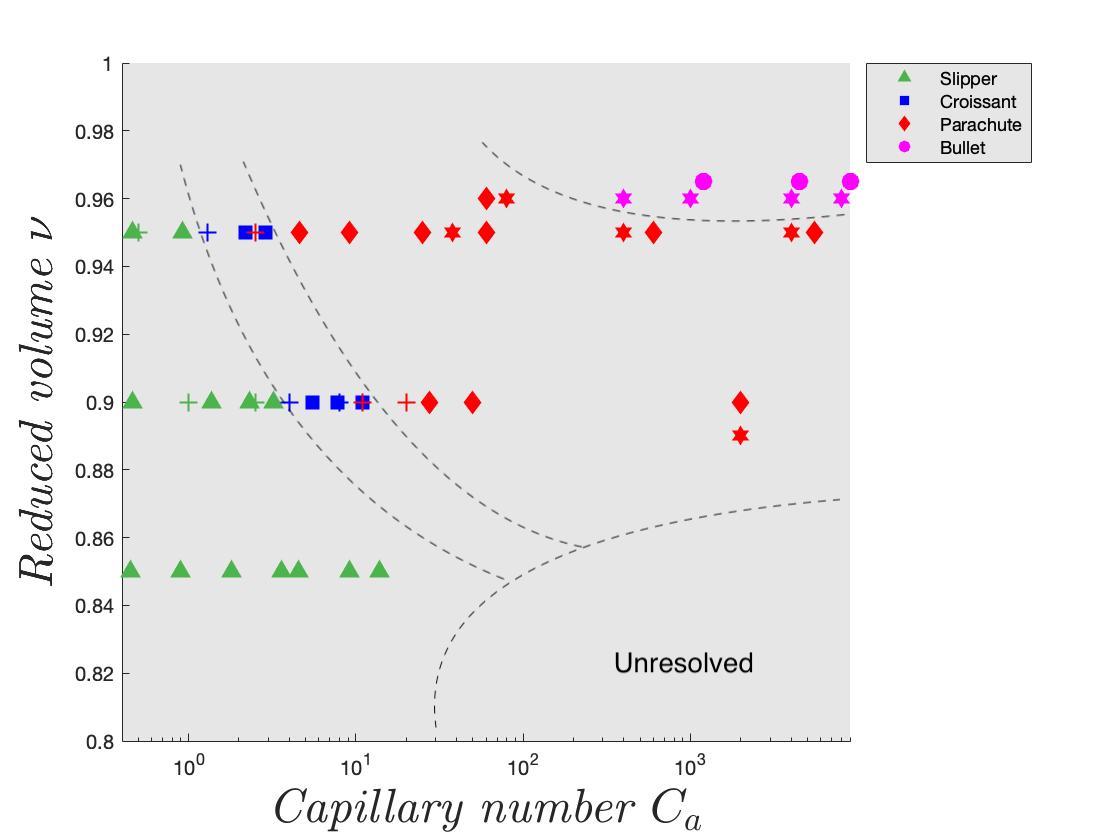}}}
     \subfloat[ Scaled equilibrium lateral positions vs capillary number for unconfined flow  \label{fig:res_lat_uncon}]{{\includegraphics[width=8.6 cm, height=7cm]{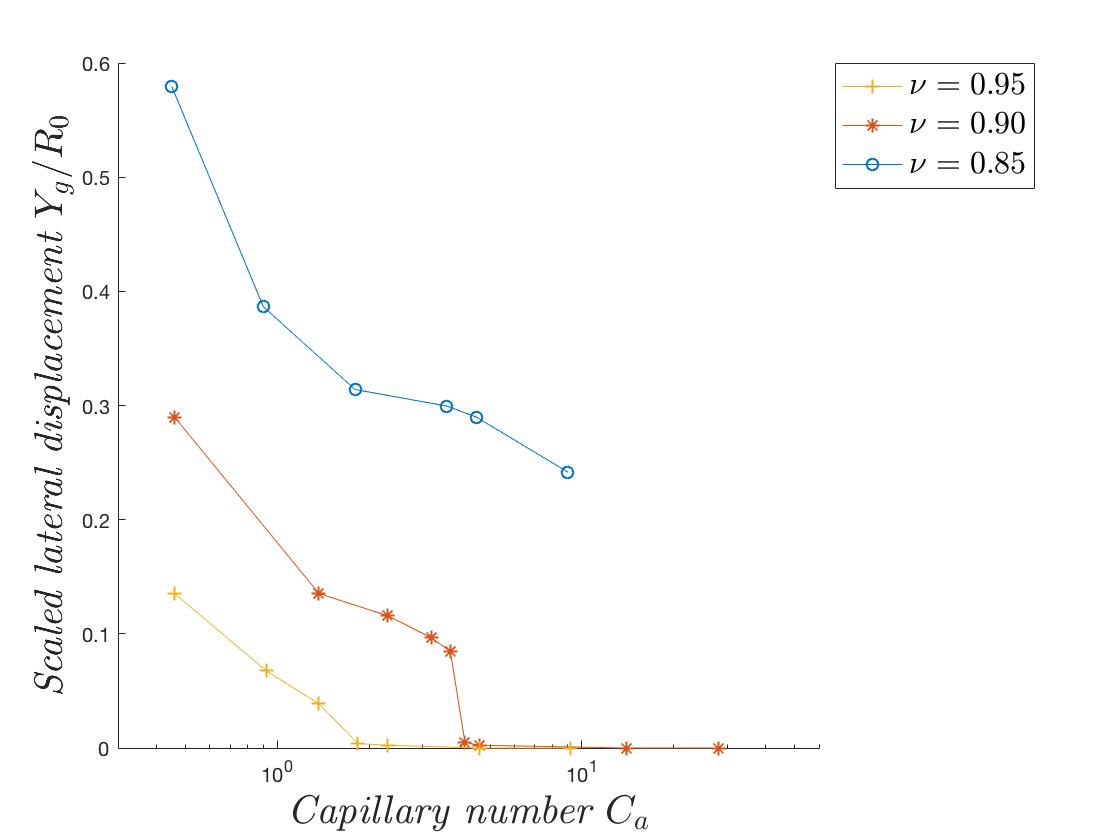}}}
     \caption{Phase diagram and equilibrium lateral positions for unconfined flow with $\lambda=1$.}
\end{figure*}

We use this study of unconfined flow with no viscosity contrast (i.e., $\lambda=1$) as a validation of our code as these results have been reported in \cite{8}. In the unconfined Poiseuille flow simulations for reduced volumes $\nu = 0.90$ and $0.95$,  ``\emph{slipper}", ``\emph{croissant}" and ``\emph{parachute}" shapes are observed as $C_{a}$ is increased. Slippers (Fig. \ref{fig:slipper_test})  are  asymmetric, off-centered and exhibit tank treading motion. Croissants (Fig. \ref{fig:croissant_test}) are slightly off-centered and semi-axisymmetric while parachutes (Fig. \ref{fig:parachute_test}) are centered and fully axisymmetric.  In particular, for $\nu = 0.95$, slippers are observed in the range $0.46 \leq C_{a} \leq 1.4$, croissants are observed in the range $1.4 < C_{a} \leq 4$ and parachutes are observed in the range $C_{a}>4$. For $\nu=0.90$, slippers are observed in the range $0.46 \leq C_{a} \leq 4$, croissants are observed in the range $4 < C_{a} \leq 15$ and parachutes are observed in the range $C_{a}>15$.  We observe that decreasing the reduced volume causes the transition from slipper to croissant and croissant to parachute to occur at higher $C_{a}$. For $\nu = 0.85$, our code was able to resolve shapes for $C_{a} \leq 10$ and we observed only slipper shapes. Simulating 3D vesicle of reduced volume $\nu<0.90$ in Poiseuille flow is a hard problem and we are not aware of any study that provides their full dynamics as a function of $C_{a}$. We combine all these results to plot a phase diagram for unconfined flow in parameter space of reduced volume ($\nu$) and capillary number ($C_{a}$), shown in Fig. \ref{fig:phase_uncon}. The scaled equilibrium lateral position of vesicle center ($Y_{g}/ R_{0}$) for different reduced volumes plotted against $C_{a}$, are presented in Fig. \ref{fig:res_lat_uncon}. We note that $Y_{g}$ reduces with increasing $C_{a}$ till it becomes zero and stays there afterwards for $\nu \geq 0.90$. For $\nu=0.85$, same behavior is observed for $Y_{g}$. Our results are in line with the analytical studies \cite{32, 33}, numerical studies \cite{8, 31} and the available experiments \cite{28, 34}. In particular, the simulation results in \cite{8} and the experimental results in \cite{28} are also plotted for comparison in Fig. \ref{fig:phase_uncon} after proper scaling of capillary numbers. The slight quantitative difference in the our results compared to the simulations in \cite{8} could be due to the difference in the radius of the Poiseuille flow which is not mentioned in \cite{8}.

\begin{figure*}
    \centering
    \subfloat[ $\lambda = 1$]{%
    \includegraphics[width=5.8cm, height=4.6cm]{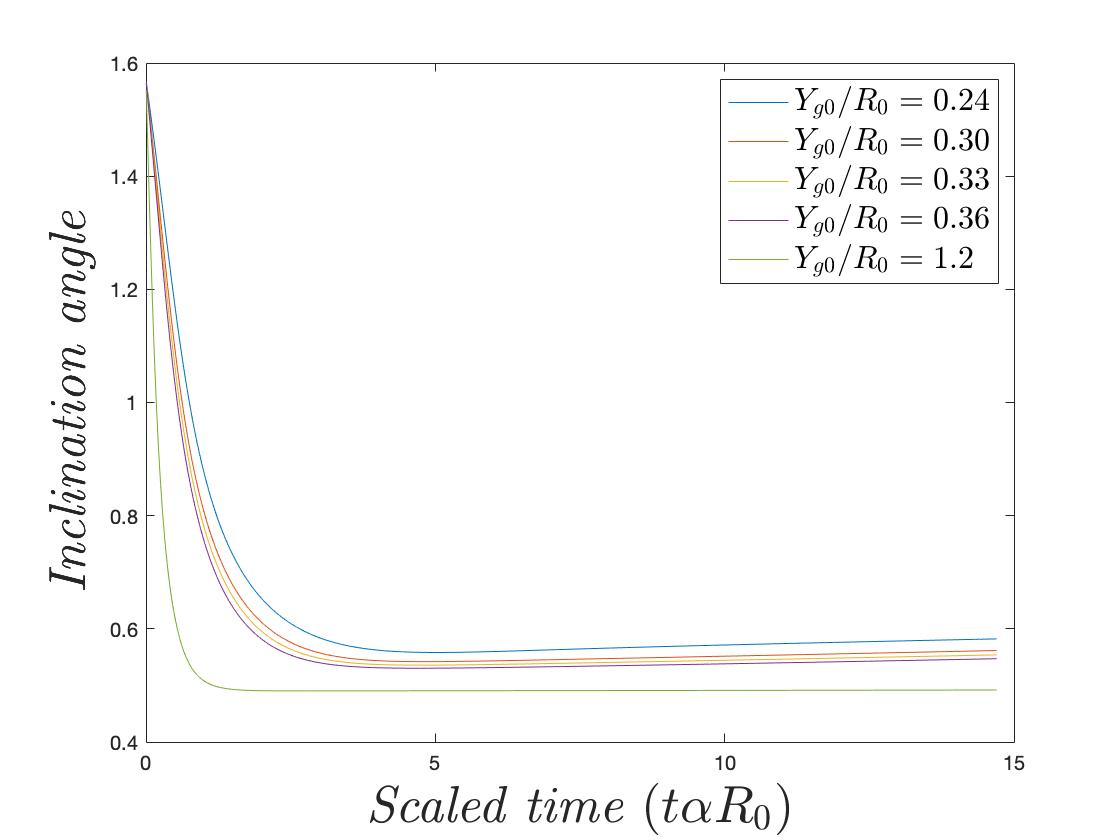} }%
    \subfloat[$\lambda = 2$]{%
    \includegraphics[width=5.8cm, height=4.6cm]{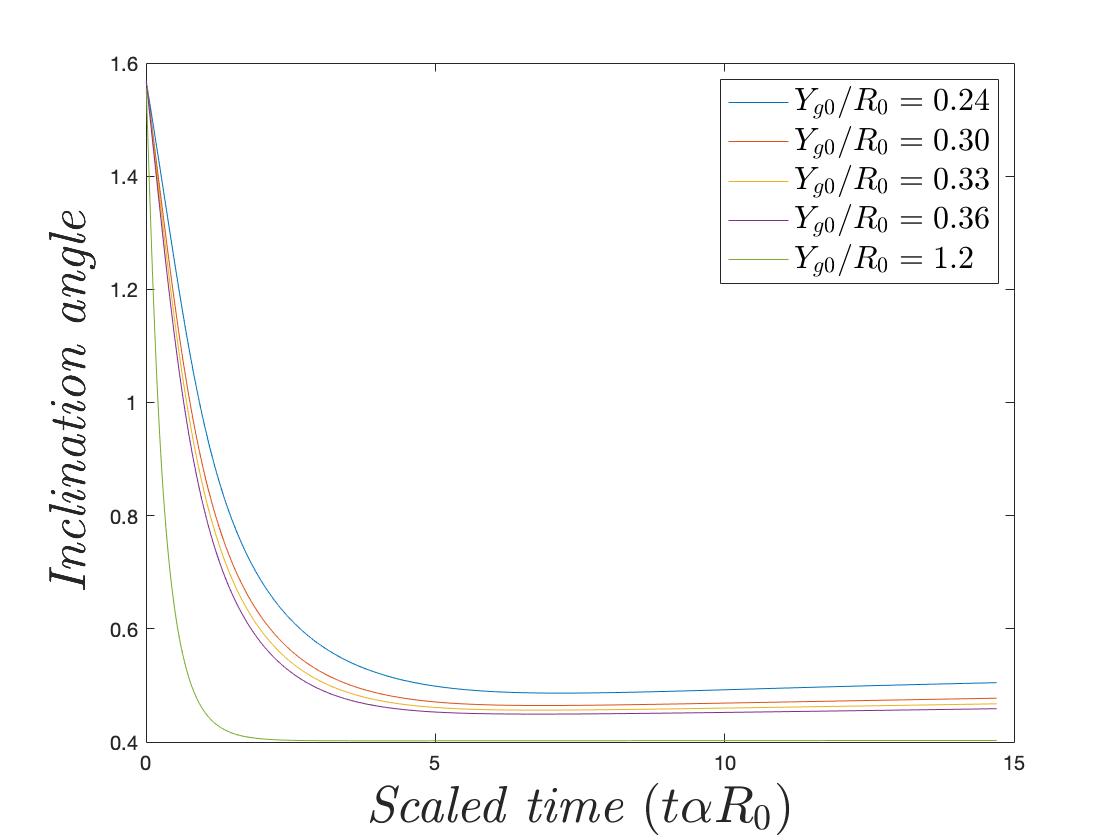} }%
    \subfloat[$\lambda = 5$ ]{%
    \includegraphics[width=5.8cm,  height=4.6cm]{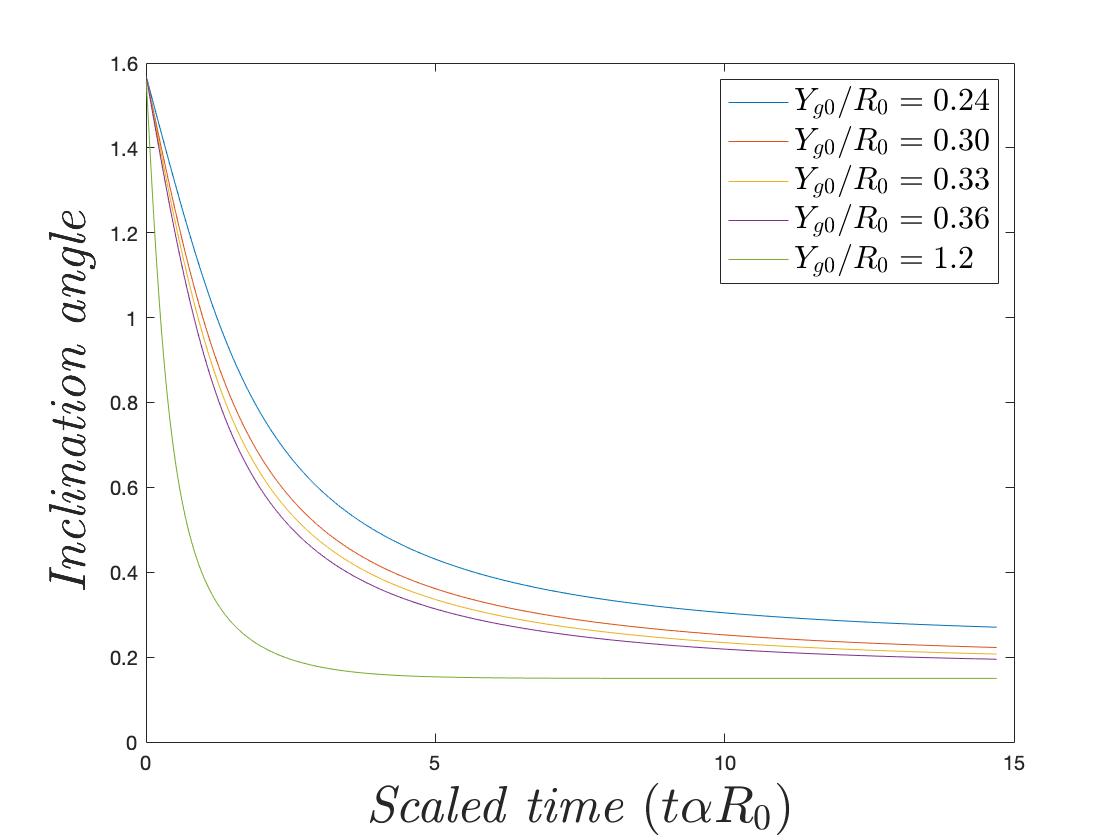} }%
    \caption{Plot of vesicle transient inclination angle (in radians; w.r.t flow direction) while migration for different initial positions in unconfined flow at $C_{a} = 3.2$. Higher viscosity contrast and higher initial lateral positions lead to lower inclination angles. Outward migration is correlated to lower inclination angles. }%
    \label{fig:migration_inclination}%
\end{figure*}

\begin{figure}
    \centering
    \includegraphics[width=9.8 cm, height=6cm]{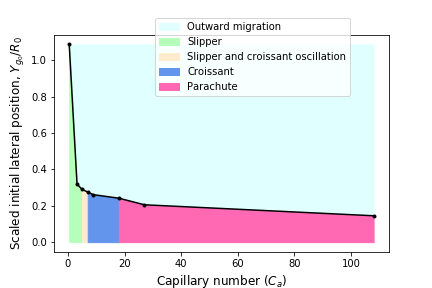}
    \caption{Unconfined flow phase diagram for vesicles of reduced volume $\nu=0.90$ with $\lambda=5$ in the parameter space of initial vesicle position ($Y_{g_{0}}$) and capillary number ($C_{a}$). For viscosity contrast $\lambda=5$, the vesicle dynamics also depend on the initial position of the vesicle. }
    \label{fig:outmig}
\end{figure}

\begin{figure}%
    \centering
    \subfloat[ $C_{n} = 0.3 $\label{fig:cn1_demo}]{{\includegraphics[width=2.8cm]{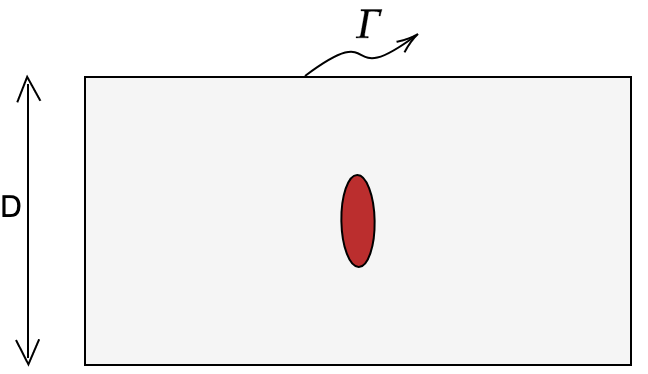} } }%
    \subfloat[$C_{n}=0.5$ \label{fig:cn2_demo}]{{\includegraphics[width=2.8cm]{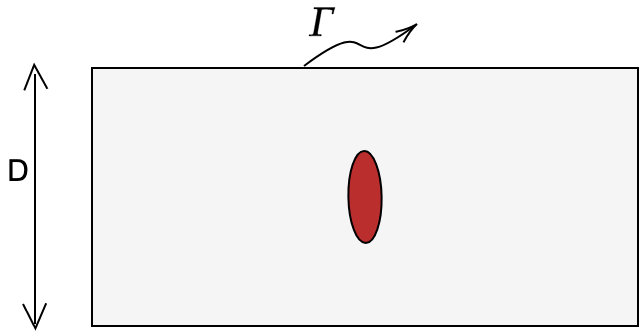} }}%
    \subfloat[$C_{n}=0.7$\label{fig:cn3_demo}]{{\includegraphics[width=2.8cm]{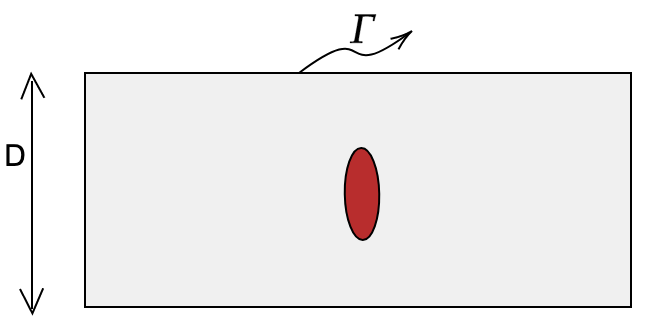} }}%
    \caption{Different confinement ratios ($C_{n}$). $\Gamma$ is the channel with diameter $D$. The vesicle is denoted in red. High confinement ratio means the bounding walls are closer to the vesicle.}%
    \label{fig:cn}%
\end{figure}

\subsection{Viscosity contrast $\lambda=5$}
When we set the viscosity contrast to $\lambda=5$, some important differences in dynamics are observed depending on the vesicle initial position. We denote vesicle center's initial distance from the centerline of the flow by $Y_{g_{0}}$. When vesicle starts close to center (i.e, $Y_{g_{0}} = 0.025R_{0}$) for $\nu=0.90$ and  $\lambda=5$, slippers, croissants and parachutes are observed as $C_{a}$ is increased but with different ranges compared to $\lambda=1$ case. We observe slippers for $0.46 \leq C_{a} \leq 5$, croissants  for $7 \leq C_{a} < 18$ and parachutes for $C_{a} \geq 18$. These observations suggest that increasing the viscosity contrast causes the shape transitions to occur at higher $C_{a}$. Similar observation was made in \cite{10} using 2D vesicle simulations. In the transition phase from slipper to croissant, for example at $C_{a}=6$,  we found a special regime in which the vesicle oscillates between the slipper and croissant shapes. something that doesn't happen in flows without viscosity contrast ($\lambda=1$). 

When the initial position of the vesicle is chosen far from the centerline $(\text{say}, Y_{g_{0}} = 1.5R_{0})$, an outward migration is observed for $0.46 \leq C_{a} \leq 1350$. The observations agree with the results in \cite{8} which reported this outward migration tendency due to the viscosity contrast.  At fixed viscosity contrast, the critical value of $Y_{g_{0}}$, above which outward migration is observed, depends on both capillary number and reduced volume. The higher the capillary number is, the lower is the critical initial position above which outward migration occurs. For example, for $\nu=0.90$ and $C_{a}=0.46$, this critical value is observed to be $1.08R_{0}$ while for $\nu=0.90$ and $C_{a}=1350$ outward migration occurs for $Y_{g_{0}} > 0.05R_{0}$ (see Fig. \ref{fig:outmig} for a complete picture of $\nu=0.90$ with $\lambda=5$). Also, the higher the reduced volume is, the higher is the critical initial position for outward migration. For example, when $\nu=0.95$ and $C_{a}=1350$, this critical initial vesicle position is observed to be $0.5R_{0}$. For $\nu=0.85$, we expected a lower value of critical initial position but, surprisingly, we observe outward migration even for very low value of $Y_{g_{0}} = 0.025R_{0}$ at all $C_{a} \geq 0.46$. We speculate that this happens because the equilibrium slipper positions of  vesicles of reduced volume $\nu=0.85$ are so high that they exceed the critical initial position for outward migration even at low $C_{a}$ (see the equilibrium positions in $\lambda=1$ case shown in Fig. \ref{fig:res_lat_uncon}). Thus, causing an outward migration even at very low capillary numbers. 

Although the exact reason for outward migration tendency remains unclear, our simulations reveal that the transient inclination angles (with respect to flow direction) of migrating vesicle can differentiate outward vs inward migration. The dependence of lift on the orientation of vesicle with respect to flow has also been discussed before in \cite{36}. In our simulations, outward migration is associated with lower values of inclination angles while inward migration is associated to higher values of inclination angles. Higher viscosity contrast leads to lower inclination angles and, thus, a strong outward migration tendency (see Fig. \ref{fig:migration_inclination}). The figure also indicates that starting far away from the center leads to lower inclination angles; thus, an increased outwards migration tendency. A further study in the direction of calculating normal stress difference as in \cite{35} could shed further light on this phenomena.

%
 \begin{figure}[]%
    \centering   
    \subfloat[ Side view]{{\includegraphics[width=3.3cm]{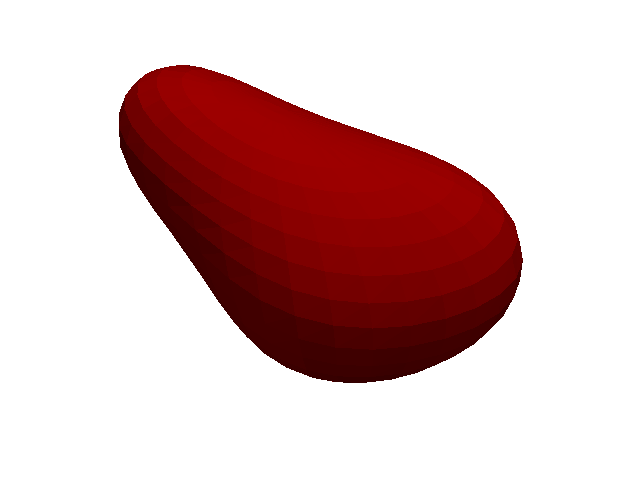} }} \hspace{1.3cm}
    \subfloat[Rear view]{{\includegraphics[width=2.9cm]{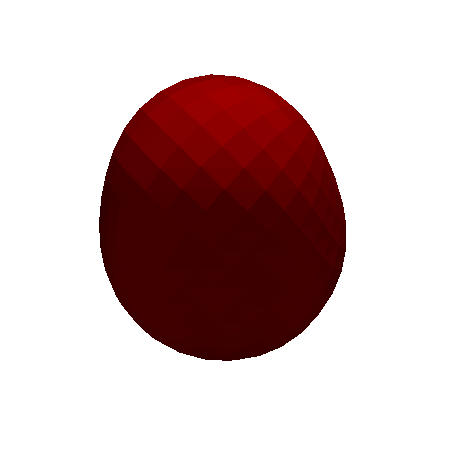} }}
    \caption{ Confined flow : Slipper  shape with $\nu =0.90$, $\lambda=1$, $C_{n}=0.3$, $C_{a}=1.5$. }
     \label{fig:confined_slipper_shape}
\end{figure}

 \begin{figure}[]%
    \centering   
    \subfloat[ Side view]{{\includegraphics[width=3.1cm]{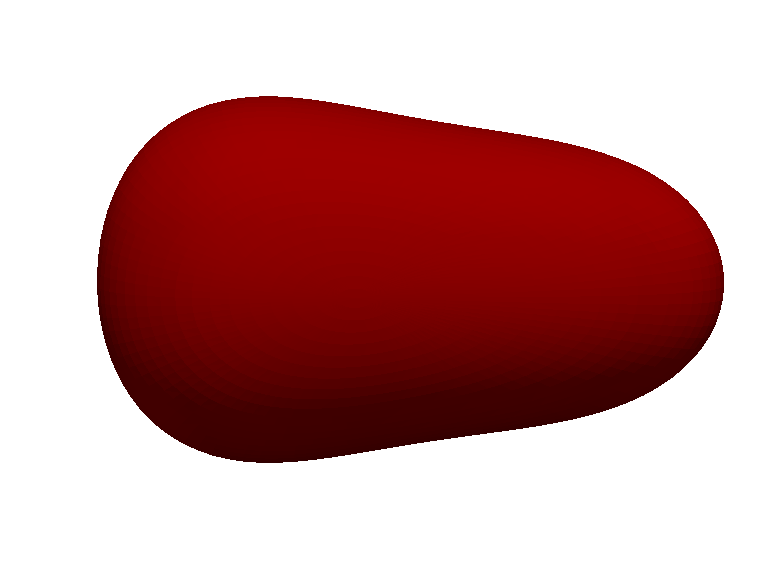} }} \hspace{1.1cm}
    \subfloat[Rear view]{{\includegraphics[width=3.1cm]{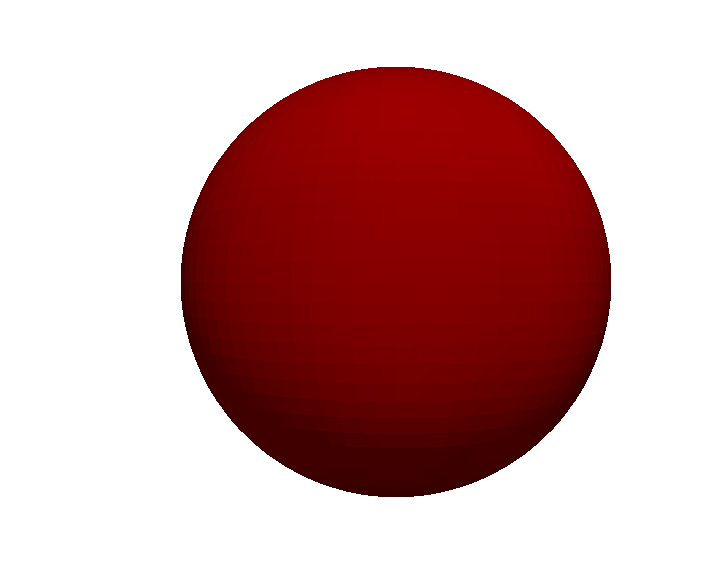} }} 
    \caption{Confined flow: Bell shape with $\nu =0.90, \lambda=1, C_{n}=0.5, C_{a} = 0.5$. }
    \label{fig:bell}
    \end{figure}
    
\begin{figure*}%
    \centering
    \subfloat[ Vesicle center ($Y_{g}$) scaled with $R_{0}$]{{\includegraphics[width=5.8cm, height=4.1cm]{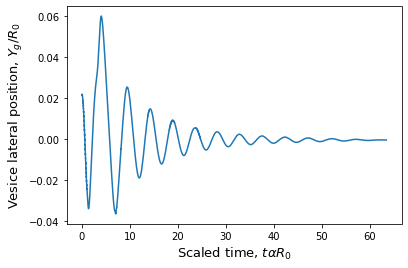} }}
    \subfloat[Bending energy of vesicle with time ]{{\includegraphics[width=5.8cm, height=4.6cm]{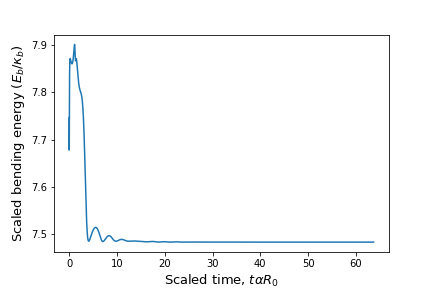} }}
    \subfloat[Vesicle moment about flow axis with time ]{{\includegraphics[width=5.8cm,  height=4.6cm]{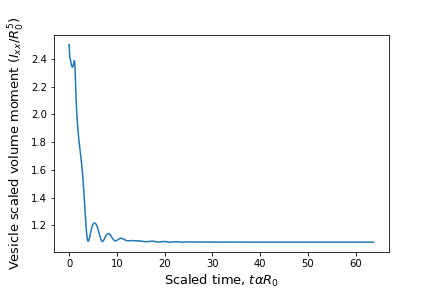} }}
    \caption{Vesicle dynamics for confined flow simulation with $\nu =0.90, \lambda=1, C_{a}=0.5, C_{n}=0.5$. Equilibrium shape is a bell. Snaking oscillations die down eventually.}%
    \label{fig:confined_bell}%
\end{figure*}

\begin{figure*}%
    \centering
    \subfloat[ Vesicle center ($Y_{g}$) scaled with $R_{0}$. Vesicle initial position $Y_{g_{0}}=0.022R_{0}$.]{{\includegraphics[width=6.8cm, height=4.1cm]{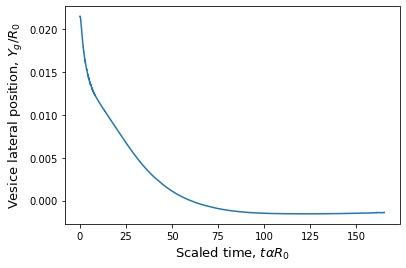} }}\hspace{2cm}
    \subfloat[ Vesicle center ($Y_{g}$) scaled with $R_{0}$. Vesicle initial position $Y_{g_{0}}=0.85R_{0}$.]{{\includegraphics[width=6.8cm, height=4.1cm]{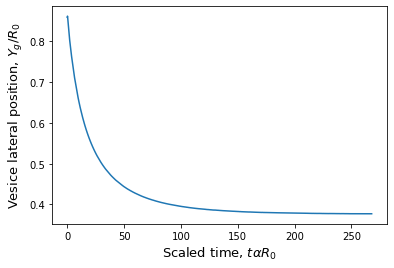} }}
    \caption{Vesicle lateral position for confined flow simulation with $\nu =0.90, \lambda=5, C_{a}=682, C_{n}=0.3$. Coexistence of slipper and parachute. a) Vesicle starts close to center. Equilibrium shape is a centered parachute. b) Vesicle starts far from center. Equilibrium lateral distance is non-zero and equilibrium shape is a slipper.}%
    \label{fig:coexistence}%
\end{figure*}

    \begin{figure}[t]
    \subfloat[ Side view]{{\includegraphics[width=2.8cm]{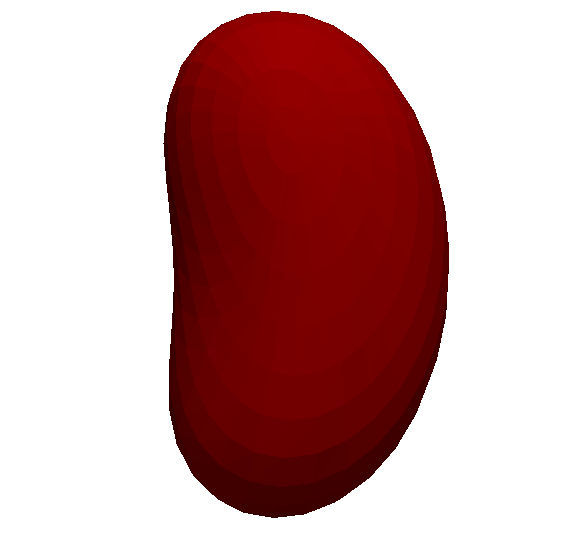} }} \hspace{1.2cm}
    \subfloat[Rear view]{{\includegraphics[width=2.6cm]{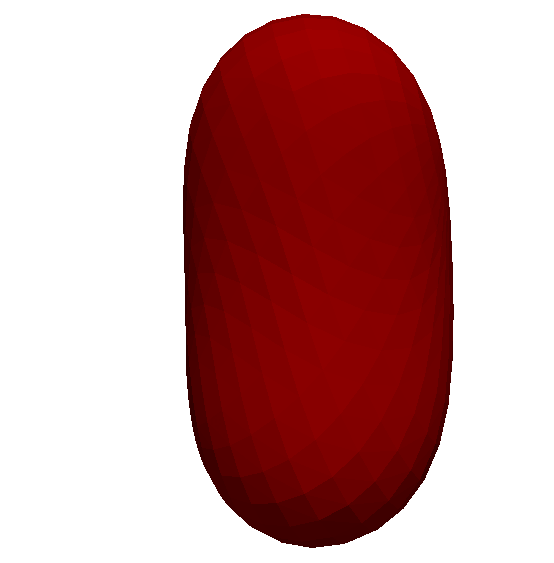} }}
    \caption{ Confined flow : Bean  shape with $\nu =0.90$, $\lambda=5$, $C_{n}=0.5$, $C_{a}=0.27$. }
    \label{fig:bean}

\end{figure}

\subsection{Discussion of results for unconfined flow}
The results for $\lambda=1$ are fairly straightforward. The equilibrium lateral position keeps on decreasing with increasing $C_{a}$ and finally becomes zero. For $\lambda=5$, the initial position changes the dynamics and it is evident that there is an outward migration tendency, which opposes the inward migration due to the quadratic component of the Poiseuille flow. When the vesicle initial position is far enough from the centerline, this outward migration tendency dominates and causes an overall outward migration velocity. Also, the lower the reduced volume is, the more dominant this outward migration seems to be.  The question is, do these observations carry to the confined flow case? This is what we try to answer in the next section.

\begin{figure*}
    \centering
    \subfloat[ A phase diagram for confined flow with $\lambda=1$.  \label{fig:phase_con1}]{{\includegraphics[width=9.8 cm, height=7cm]{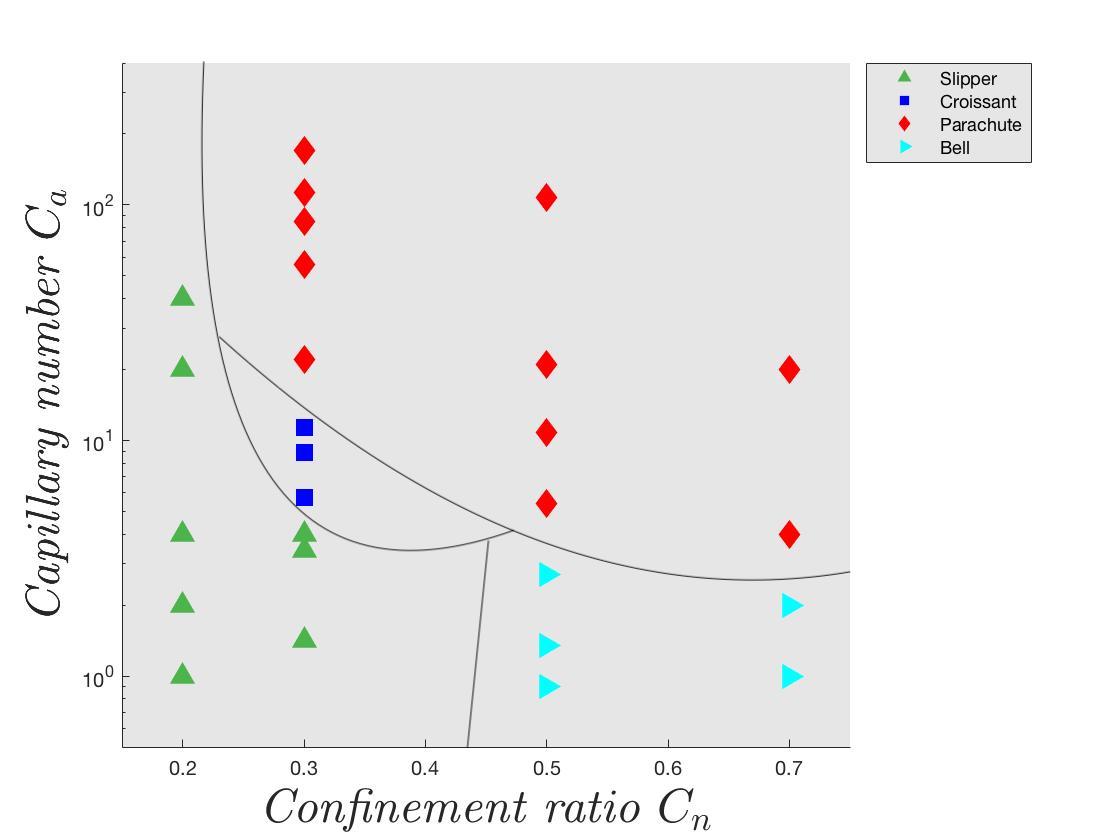}}}
     \subfloat[ A phase diagram for confined flow with $\lambda=5$.  \label{fig:phase_con5}]{{\includegraphics[width=9.8 cm, height=7cm]{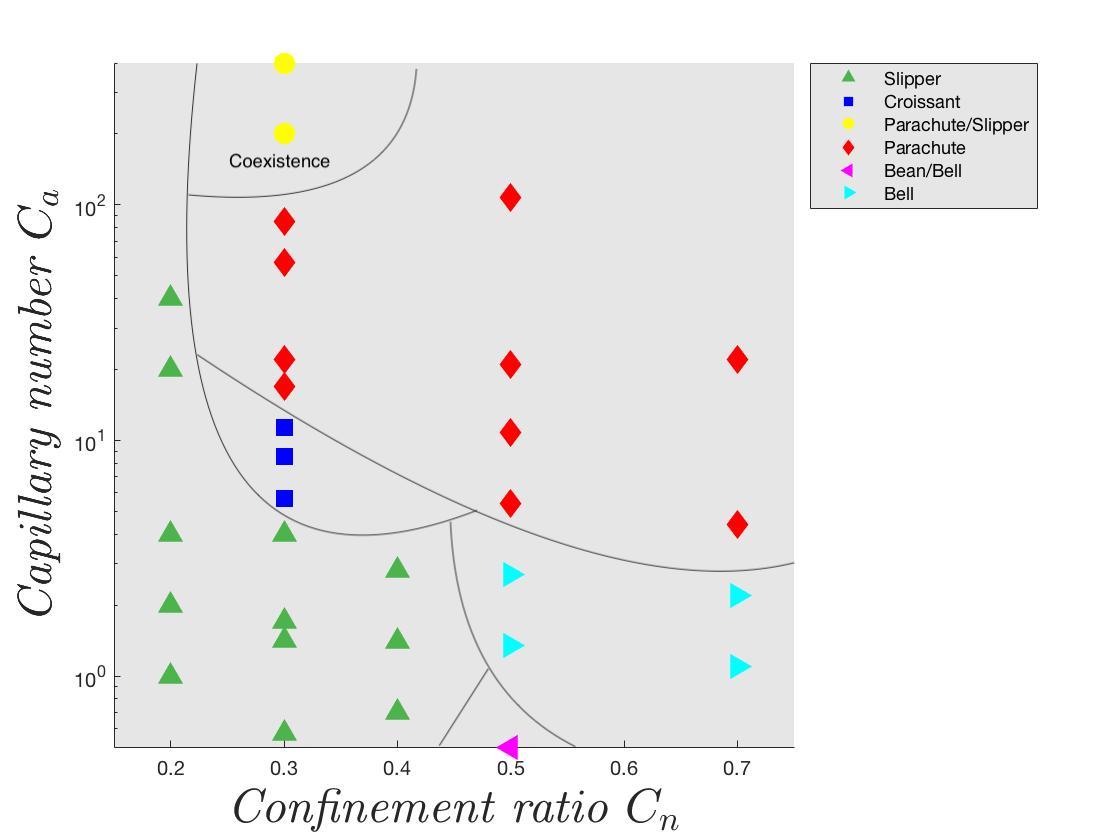}}}
     \caption{Phase diagrams for confined flow in the parameter space of confinement ratio  $C_{n}$ and capillary number $C_{a}$. Black curves are a guide to the eye.}
	\label{fig:phase_con}
\end{figure*}

\section{\label{sec:phaseconfined}Results for confined flow}
In this section, we study the effect of wall confinement (see Fig. \ref{fig:cn}) on vesicle dynamics. We first study the case with no viscosity contrast for different confinement ratios. Then we proceed onto the case with viscosity contrast. We use vesicle of reduced volume $\nu=0.90$ for this study and compare it with unconfined flow dynamics.

\subsection{Viscosity contrast $\lambda =1$}
First, we consider the case of low confinement ratio, i.e., $C_{n} = 0.3$. In this case, it is natural to expect that the difference between the confined and the unconfined Poiseuille flow would be negligible due to the weak hydrodynamic effect of the walls. Hence, we should expect the same qualitative behavior as in the unconfined case. The simulations indeed confirm our expectation. We observe slippers (Fig. \ref{fig:confined_slipper_shape}) in the range $1 \leq C_{a} \leq 4.5$ with equilibrium lateral position decreasing as $C_{a}$ increases. In the range $4.5 \leq C_{a}  \leq 13$, croissants are observed and parachutes are observed for $C_{a}> 13$. 

But, increasing the confinement ratio to $0.5$ paints a different picture.  A centered bell shape (see Fig. \ref{fig:bell}) is observed for $0.134 \leq C_{a} \leq 3$. The bell shape is axisymmetric like the parachute shape but the rear is convex, i.e., bulges outwards instead of the concave rear in the parachute shape. Note that initially the bell seems to exhibit \emph{``snaking"} motion but after some time the motion dies and a centered stationary bell is observed (see Fig. \ref{fig:confined_bell}).  For $C_{a} > 3$, the rear of the shape becomes concave and we get a parachute. On further increasing the confinement ratio to $C_{n}=0.7$, a centered bell shape is observed in the range $0.5 \leq C_{a} \leq 2.5$ and a parachute shape is observed for $C_{a} > 2.5$. The occurrence of axisymmetric centered shapes at low $C_{a}$ suggests a dominance of the effects of confining walls which push the vesicle towards the centerline. 

\subsection{Viscosity contrast $\lambda=5$}
Based on our earlier results,  we expected this to be the most interesting case since the inward push from the confining walls (for details on wall push, refer to \cite{34, 36, 37} ) and the outward migration tendency (due to the viscosity contrast) oppose each other and can result in interesting dynamics. And indeed that is the case. For low confinement ratio $C_{n}=0.3$, if the vesicle initial position is close to centerline ($Y_{g_{0}}\leq0.04R_{0}$), we observe a slipper shape for $0.05 \leq C_{a} \leq 5$, a croissant for $5 \leq C_{a} \leq 10$ and a parachute for $10 \leq C_{a}  \leq 680$ similar to the no contrast case. But, interestingly, for $C_{a} \geq 100$, a bistability  is observed, i.e., slipper and parachute equilibrium shapes coexist depending on initial position of the vesicle. For example, for $C_{a}=682$, if vesicle starts far away from the centerline ($Y_{g_{0}} > 0.30R_{0}$), the equilibrium shape is a tank treading slipper at mean position $Y_{g} = 0.38R_{0}$ while if it starts close to center, the equilibrium shape is a centered parachute (see Fig. \ref{fig:coexistence}). To ascertain if this coexistence occurs because of the cancellation of outward migration tendency by the confinement effects, we repeat the simulation with initial position slightly less than the equilibrium slipper position but without confining walls. We observe that for the same parameters without confining walls, the vesicle continued to migrate outwards perpetually while it stopped at an equilibrium position when confining walls are present. This confirms that this bistability is a result of the cancellation of outward migration tendency by the confinement effects. We have seen that outward migration is strong at high $C_{a}$ in unconfined flow simulations which could explain why no such bistability is observed for $C_{a}<100$. 

For $C_{n}=0.5$, the confinement effects dominate as observed in the no viscosity contrast case. Although we do not observe coexistence of slipper and parachute in this case, we do observe a coexistence of bean (see Fig. \ref{fig:bean}) and bell shape for $0.1 \leq C_{a} \leq 0.4$ depending on the initial position. The reason for this bistability remains unclear. For $0.4 < C_{a} < 4$, only bell shape is observed while parachute is observed for $C_{a} \geq 4$. For $C_{n}=0.7$, bell and parachute shapes are observed as before when $C_{a}$ is increased.

\subsection{Discussion of results for confined flow}
We combine the above results for confined flow without and with viscosity contrast in parameter space of capillary number and confinement ratio to plot the phase diagrams  in Fig. \ref{fig:phase_con}. We conclude that for low confinement ratio, the dynamics is largely similar to the unconfined flow. Slipper shapes are common in low confinements although parachutes are also observed at high flow strength. Interestingly, the presence of viscosity contrast induces a bistability at high flow strength, i.e., both slippers and parachutes are observed. At higher confinement ratios ($C_{n} \geq 0.5$), confinement effects dominate the dynamics and axisymmetric shapes (bell and parachute) are observed. While with $\nu = 0.90$, we do not observe coexistence of slipper and parachute in higher confinements, we should keep in mind that lower reduced volume vesicles might exhibit this coexistence since the outward migration tendency seems to become stronger as reduced volume is decreased. This could be a possible explanation for the experimental observation of slipper shapes of RBCs in microcapillaries at high velocities which have reduced volume of about 0.7 and viscosity contrast about five \cite{29,30}.

\section{\label{sec:conc} Conclusion and Future Work }

In this paper, we have used vesicles as a model of RBCs to provide a picture of their dynamics and equilibrium shapes  in confined Poiseuille flow with and without viscosity contrast. The phase diagrams for both the cases have been provided. To our knowledge, this is the first study that provides a phase diagram for 3D vesicles with viscosity contrast in confined Poiseuille flow. We have seen how the parabolic velocity profile of Poiseuille flow, viscosity contrast and the confining walls affect the dynamics and shapes of vesicles for a range of relevant parameters. Although in the case with no viscosity contrast, we have seen that slipper shape occurs on decreasing the flow strength, the experiments with RBCs point to the opposite. Our results on the bistability created due to the presence of viscosity contrast and the confining walls could explain this anomaly. But unlike RBCs, vesicles have no shear resistance which leads to large deformations. It should be interesting to study how the dynamics of capsules (which have shear resistance) compare with vesicles in both unconfined and confined Poiseuille flow. This is a future direction that we aim to explore in our future work.


\begin{thebibliography}{9}
\bibitem{1}
 F. Merola, P. Memmolo, L. Miccio, R. Savoia, M. Mugnano,
A. Fontana, G. D'Ippolito, A. Sardo, A. Iolascon,
A. Gambale, and P. Ferraro, \textit{``Tomographic Flow Cytometry
by Digital Holography," } Light Sci. Appl. \textbf{6}, e16241
(2017)

\bibitem{2}
V. Vitkova, M.-A. Mader, B. Polack, C. Misbah, and
T. Podgorski, \textit{``Micro-Macro Link in Rheology of Erythrocyte
and Vesicle Suspensions,"} Biophys. J. \textbf{95}, L33
(2008).

\bibitem{3}
 E. Henry, S. H. Holm, Z. Zhang, J. P. Beech, J. O. Tegenfeldt,
D. A. Fedosov, and G. Gompper, \textit{``Sorting cells by
their dynamical properties,"} Sci. Rep. \textbf{6}, 34375 (2016)


\bibitem{5}
 P. Gaehtgens, C. D\"{u}hrssen, and K. H. Albrecht, \textit{``Motion,
deformation, and interaction of blood cells and plasma
during flow through narrow capillary tubes"}, Blood Cells
\textbf{6}, 799 (1980)

\bibitem{6}
B. Kaoui, G. Biros, and C. Misbah, \textit{``Why Do Red Blood Cells Have Asymmetric Shapes Even in a Symmetric Flow?"}, Physical Review Letters \textbf{103}, 188101 (2009)

\bibitem{7}
B. Kaoui, N. Tahiri, T. Biben, H. Ez-Zahraouy, A. Benyoussef, G. Biros, and C. Misbah, \textit{``Complexity of vesicle microcirculation"}, Physical Review E \textbf{84}, 041906 (2011)

\bibitem{10}
 N. Tahiri , T. Biben , H. Ez-Zahraouy , A. Benyoussef , C. Misbah \textit{``On the problem of slipper shapes of red blood cells in the microvasculature"}, Microvascular Research \textbf{85}, 40 (2013)

\bibitem{8}
 A. Farutin and C. Misbah, \textit{``Symmetry breaking and cross-streamline migration of three-dimensional
vesicles in an axial Poiseuille flow"}, Physical Review E  \textbf{89}, 042709 (2014)


\bibitem{28}
G. Coupier,  A. Farutin, C. Minetti, T. Podgorski,  C. Misbah, \textit{``Shape diagram of vesicles in Poiseuille 
flow"}, Phys. Rev. Lett. \textbf{108}, 178106 (2012)





\bibitem{29}
G. Tomaiuolo, M. Simeone, V. Martinelli, B. Rotoli, and
S. Guido, \textit{``Red blood cell deformation in microconfined flow"},Soft Matter \textbf{5}, 3736 (2009)

\bibitem{30}
S. Guido and G. Tomaiuolo, \textit{``Microconfined flow behavior of red blood cells in vitro "}C. R. Phys. \textbf{10}, 752 (2009).

\bibitem{21}
C. Pozrikidis, \textit{``Boundary Integral and Singularity Methods for Linearized Viscous Flow"}, Cambridge
University Press, Cambridge, (1992)

\bibitem{22}
D. Malhotra, A. Rahimian, D. Zorin, G. Biros, \textit{``A parallel algorithm for long-timescale simulation of concentrated
vesicle suspensions in three dimensions"}, (2016) (unpublished)

\bibitem{23}
C. Pozrikidis,  \textit{``Effect of membrane bending stiffness on the deformation of capsules in simple
shear flow"}, Journal of Fluid Mechanics \textbf{440}, 269 (2001)

\bibitem{24}
J. L. Weiner, \textit{``On a problem of Chen, Willmore, et al."}, Indiana University Mathematics Journal 
\textbf{27}, 19 (1978)

\bibitem{25}
C. Pozrikidis, \textit{``Interfacial dynamics for Stokes flow. Journal of Computational Physics"}, \textbf{169}, 250 (2001)

\bibitem{26}
A. Rahimian, S. Veerapaneni, G. Biros , \textit{``Dynamic simulation of locally inextensible vesicles suspended in an
arbitrary two-dimensional domain, a boundary integral method"},  Journal of Computational Physics \textbf{229}, 6466 (2010)

\bibitem{20}
A. Farutin, T. Biben, C. Misbah, \textit{``3D numerical simulations of vesicle and inextensible capsule dynamics"}, Journal of Computational Physics \textbf{275}, 539 (2014) 



\bibitem{32}
G. Danker, P. M. Vlahovska, and C. Misbah, \textit{``Vesicles in Poiseuille Flow"}, Phys. Rev. Lett.
\textbf{102}, 148102 (2009).

\bibitem{33}
A. Farutin, T. Biben, and C. Misbah, {``Analytical progress in the theory of vesicles under linear flow"}, Phys. Rev. E \textbf{81}, 061904
(2010).


\bibitem{31}
B. Kaoui, G. H. Ristow, I. Cantat, C. Misbah, and W.
Zimmermann, \textit{``Lateral migration of a two-dimensional vesicle in unbounded Poiseuille flow"}, Phys. Rev. E \textbf{77}, 021903 (2008).

\bibitem{34}
G. Coupier, B. Kaoui, T. Podgorski, and C.Misbah, \textit{``Non-inertial lateral migration of vesicles in bounded Poiseuille flow"}, Phys. Fluids
\textbf{20}, 111702 (2008).


\bibitem{35}
G. Ghigliotti, A. Rahimian, G. Biros, and C. Misbah, \textit{``Vesicle Migration and Spatial Organization Driven by Flow Line Curvature"}, Phys. Rev. Lett. \textbf{106}, 028101 (2011).


\bibitem{36}
P. Olla, \textit{``The lift on a tank-treading ellipsoidal cell in a shear flow"}, J. Phys. II France \textbf{7}, 1533 (1997).


\bibitem{37}
I. Cantat and C. Misbah, \textit{``Lift force and dynamical unbinding of adhering vesicles under shear flow"}, Phys. Rev. Lett. \textbf{83}, 880 (1999).

%
%
%
%

%
%


%


















\end{thebibliography}
\end{document}